\documentclass[10pt, a4paper, oneside, headinclude, footinclude]{scrartcl}

\usepackage[
nochapters, % Turn off chapters since this is an article        
beramono, % Use the Bera Mono font for monospaced text (\texttt)
pdfspacing, % Makes use of pdftex’ letter spacing capabilities via the microtype package
dottedtoc % Dotted lines leading to the page numbers in the table of contents
]{classicthesis} % The layout is based on the Classic Thesis style

\usepackage{arsclassica} % Modifies the Classic Thesis package

\usepackage[T1]{fontenc} % Use 8-bit encoding that has 256 glyphs

\usepackage[utf8]{inputenc} % Required for including letters with accents

\usepackage{graphicx} % Required for including images
\graphicspath{{Figures/}} % Set the default folder for images

\usepackage{enumitem} % Required for manipulating the whitespace between and within lists

\usepackage{lipsum} % Used for inserting dummy 'Lorem ipsum' text into the template

\usepackage{subcaption} % Required for creating figures with multiple parts (subfigures)

\usepackage{amsmath,amssymb,amsthm} % For including math equations, theorems, symbols, etc

\theoremstyle{definition} % Define theorem styles here based on the definition style (used for definitions and examples)

\theoremstyle{plain} % Define theorem styles here based on the plain style (used for theorems, lemmas, propositions)

\theoremstyle{remark} % Define theorem styles here based on the remark style (used for remarks and notes)

\hypersetup{
colorlinks=true, breaklinks=true, bookmarks=true,bookmarksnumbered,
urlcolor=webbrown, linkcolor=RoyalBlue, citecolor=webgreen, % Link colors
pdftitle={}, % PDF title
pdfauthor={\textcopyright}, % PDF Author
pdfsubject={}, % PDF Subject
pdfkeywords={}, % PDF Keywords
pdfcreator={pdfLaTeX}, % PDF Creator
pdfproducer={LaTeX with hyperref and ClassicThesis} % PDF producer
} 

\usepackage[left = 4cm, right = 4cm, top = 2cm, bottom = 2.8cm]{geometry}
\usepackage{authblk}
\usepackage{bm}
\usepackage{centernot}
\usepackage{float}
\usepackage[sectionbib]{natbib}
\usepackage{pgfplots}
\usepackage{tabularx} 

\usetikzlibrary{arrows}
\usetikzlibrary{calc}
\usetikzlibrary{fit}
\usetikzlibrary{positioning}
\usetikzlibrary{shapes}

\bibpunct{(}{)}{;}{a}{}{;}

\allowdisplaybreaks

\title{\spacedallcaps{\large Bayesian Estimation of Economic Simulation Models using Neural Networks}}

\author[1, 2]{\spacedlowsmallcaps{Donovan Platt} \thanks{Corresponding author, donovan.platt@maths.ox.ac.uk}}

\affil[1]{\small Mathematical Institute, University of Oxford}
\affil[2]{\small Institute for New Economic Thinking (INET) at the Oxford Martin School}

\date{}

\begin{document}

%----------------------------------------------------------------------------------------
% Header
%----------------------------------------------------------------------------------------

\clearscrheadfoot

\cfoot[\pagemark]{\pagemark}

%----------------------------------------------------------------------------------------
% Create Front Page and Abstract
%----------------------------------------------------------------------------------------

\maketitle
\setcounter{tocdepth}{2}

\vspace{-1.5cm}

\section*{\centerline{Abstract}}

\begin{abstract}
\noindent Recent advances in computing power and the potential to make more realistic assumptions due to increased flexibility have led to the increased prevalence of simulation models in economics. While models of this class, and particularly agent-based models, are able to replicate a number of empirically-observed stylised facts not easily recovered by more traditional alternatives, such models remain notoriously difficult to estimate due to their lack of tractable likelihood functions. While the estimation literature continues to grow, existing attempts have approached the problem primarily from a frequentist perspective, with the Bayesian estimation literature remaining comparatively less developed. For this reason, we introduce a Bayesian estimation protocol that makes use of deep neural networks to construct an approximation to the likelihood, which we then benchmark against a prominent alternative from the existing literature. Overall, we find that our proposed methodology consistently results in more accurate estimates in a variety of settings, including the estimation of financial heterogeneous agent models and the identification of changes in dynamics occurring in models incorporating structural breaks.
\end{abstract}

\vspace{1cm}

\noindent \textbf{Keywords}: Agent-based modelling, Simulation modelling, Bayesian estimation, Machine learning, Neural networks

\vspace{0.5cm}

\noindent \textbf{JEL Classification}: C13 $\cdot$ C52

%----------------------------------------------------------------------------------------
% Article Text
%----------------------------------------------------------------------------------------

\section{Introduction}

Recent years have, to some extent, seen the emergence of a paradigm shift in how economic models are constructed. Traditionally, a need to facilitate mathematical tractability and limited computational resources have led to a dependence on strong assumptions\footnote{These include, but are not limited to, assumptions of perfect rationality and the existence of representative agents.}, many of which are inconsistent with the heterogeneity and non-linearity that characterise real economic systems \citep{Farmer_Geanakoplos_2009, Farmer_Foley_2009, Fagiolo_Roventini_2017}. Ultimately, the Great Recession of the late 2000s and the perceived failings of traditional approaches, particularly those built on general equilibrium theory, would lead to the birth of a growing community arguing that the adoption of new paradigms harnessing contemporary advances in computing power could lead to richer and more robust insights \citep{Farmer_Foley_2009, Fagiolo_Roventini_2017}.

Perhaps the most prominent examples of this new wave of computational approaches are agent-based models (ABMs), which attempt to model systems by directly simulating the actions of and interactions between their microconstituents \citep{Macal_North_2010}. In theory, the flexibility offered by simulation should allow for more empirically-motivated assumptions and this, in turn, should result in a more principled approach to the modelling of the economy \citep{Chen_2003, LeBaron_2006}. The extent to which this has been achieved in practice, however, remains open for debate \citep{Hamill_Gilbert_2016}.

While ABMs initially found success by demonstrating an ability to replicate a wide array of stylised facts not recovered by more traditional approaches \citep{LeBaron_2006, Barde_2016}, their simulation-based nature makes their estimation nontrivial \citep{Fagiolo_et_al_2017}. Therefore, while the last decade has seen the emergence of increasingly larger and more realistic macroeconomic models, such as the Eurace \citep{Cincotti_et_al_2010} and Schumpeter Meeting Keynes \citep{Dosi_et_al_2010} models, their acceptance in mainstream policy-making circles remains limited due to these and other challenges.

The aforementioned estimation difficulties largely stem from the simulation-based nature of ABMs, which, in all but a few exceptional cases\footnote{See, for example, the work of \citet{Alfarano_et_al_2005}, \citet{Alfarano_et_al_2006} and \citet{Alfarano_et_al_2007}.}, renders it impossible to obtain a tractable expression for the likelihood function. As a result, most existing approaches have attempted to circumvent these difficulties by directly comparing model-simulated and empirically-measured data using measures of dissimilarity (or similarity) and searching the parameter space for appropriate values that minimise (or maximise) these metrics \citep{Grazzini_et_al_2017, Lux_2018}. The most pervasive of these approaches, which \citet{Grazzini_Richiardi_2015} call simulated minimum distance (SMD) methods, is the method of simulated moments (MSM), which constructs an objective function by considering weighted sums of the squared errors between simulated and empirically-measured moments (or summary statistics).

Though MSM has been widely applied in a number of different contexts\footnote{See \citet{Franke_2009}, \citet{Franke_Westerhoff_2012}, \citet{Fabretti_2013}, \citet{Grazzini_Richiardi_2015}, \citet{Chen_Lux_2016} and \citet{Platt_Gebbie_2018} for examples.} and has desirable mathematical properties\footnote{The estimator is both consistent and asymptotically normal \citep{McFadden_1989}.}, it suffers from a critical weakness. In more detail, the choice of moments or summary statistics is entirely arbitrary and the quality of the associated parameter estimates depends critically on selecting a sufficiently comprehensive set of moments, which has proven to be nontrivial in practice. In response, recent years have seen the development of a new generation of SMD methods that largely eliminate the need to transform data into a set of summary statistics and instead harness its full informational content \citep{Grazzini_et_al_2017}.

These new methodologies vary substantially in their sophistication and theoretical underpinnings. Among the simplest of these approaches is attempting to match time series trajectories directly, as suggested by \citet{Recchioni_et_al_2015}. More sophisticated alternatives include information-theoretic approaches \citep{Barde_2017, Lamperti_2017}, simulated maximum likelihood estimation \citep{Kukacka_Barunik_2017}, and comparing the causal mechanisms underlying real and simulated data through the use of SVAR regressions \citep{Guerini_Moneta_2017}. In addition to the development of similarity metrics, attempts have also been made to reduce the large computational burden imposed by SMD methods by replacing the costly model simulation process with computationally efficient surrogates \citep{Salle_Yildizoglu_2014, Lamperti_et_al_2018}. 

Interestingly, the aforementioned approaches are all frequentist in nature, with Bayesian estimation being significantly less prevalent\footnote{There is a rather substantial literature on what are called approximate bayesian computation methods that has gained a significant following in biology and ecology \citep{Sisson_et_al_2018}. Unfortunately, the vast majority of these methods rely on converting data to a set of summary statistics and their appeal for estimating economic ABMs is therefore limited.}. As it currently stands, only one study in the literature \citep{Grazzini_et_al_2017} has focused extensively on the use of Bayesian techniques and recent work by \citet{Lux_2018} involving sequential Monte Carlo methods includes attempts at Bayesian estimation, though the work as a whole focuses more on a frequentist approach.

While the estimation literature has certainly been growing, it still suffers from a number of key weaknesses. Perhaps the most significant of these is a lack of a standard benchmark against which to compare the performance of new methods. For this reason, most new approaches have traditionally only been tested in isolation and comparative exercises have been relatively rare. For this reason, we compared a number of prominent estimation techniques in a previous investigation \citep{Platt_2019} and found, rather surprisingly, that the Bayesian estimation procedure proposed by \citet{Grazzini_et_al_2017} consistently outperformed a number of prominent frequentist alternatives in a series of head-to-head tests, despite its relative simplicity. We therefore argued that more interest in Bayesian methods is warranted and suggested that increased emphasis should be placed on their development.

In line with this recommendation, we introduce a method for the Bayesian estimation of economic simulation models\footnote{It is worth noting that while we focus on ABMs, the proposed methodology is applicable to any model capable of simulating time series or panel data. For this reason, the methodology would be equally applicable to competing modelling approaches.} that relaxes a number of the assumptions made by the approach of \citet{Grazzini_et_al_2017} through the use of a neural network-based likelihood approximation. We then benchmark our proposed methodology through a series of computational experiments and finally conclude with discussions related to practical considerations, such as the setting of the method's hyperparameters and the associated computational costs. 

\section{Estimation and Experimental Procedures} \label{Experimental_Procedures}

In this section, we introduce the reader to a number of the essential elements of our investigation, including a brief discussion of the fundamentals of Bayesian estimation, a description of the approach of \citet{Grazzini_et_al_2017} (our chosen benchmark), and an introduction to our proposed estimation methodology.

\subsection{Bayesian Estimation of Simulation Models}

For our purposes, we consider a simulation model to be any mathematical or algorithmic representation of a real world system capable of producing time series (panel) data of the form
\begin{equation}
\bm{X} ^{sim} (\bm{\theta}, T, i) = \left [\bm{x} ^{sim} _{1, i}(\bm{\theta}), \bm{x} ^{sim} _{2, i}(\bm{\theta}), \dots, \bm{x} ^{sim} _{T, i}(\bm{\theta}) \right],
\end{equation}
where $\bm{\theta}$ is a model parameter set in the space of feasible parameter values, $T$ is the length of the simulation, $i$ represents the seed used to initialise the model's random number generators, and $\bm{x} ^{sim} _{t, i}(\bm{\theta}) \in \mathbb{R}^{n}$ for all $t = 1, 2, \dots, T$.

In general, estimation or calibration procedures aim to determine appropriate values for $\bm{\theta}$ such that $\bm{X} ^{sim} (\bm{\theta}, T, i)$ produces dynamics that are as close as possible to those observed in an empirically-measured equivalent,
\begin{equation}
\bm{X} = \left [\bm{x}_{1}, \bm{x}_{2}, \dots, \bm{x}_{T} \right],
\end{equation} 
where $\bm{x}_{t} \in \mathbb{R}^{n}$ for all $t = 1, 2, \dots, T$.

Bayesian estimation attempts to achieve the above by first assuming that the parameter values follow a given distribution, $p(\bm{\theta})$, which is chosen to reflect one's prior knowledge or beliefs regarding the parameter values. This is then updated in light of empirically-measured data, yielding a modified distribution, $p(\bm{\theta} | \bm{X})$, called the posterior. Bayesian estimation can therefore be framed in terms of Bayes' theorem as follows:
\begin{equation}
p(\bm{\theta} | \bm{X}) = \frac{p(\bm{X} | \bm{\theta})p(\bm{\theta})}{p(\bm{X})}.
\end{equation}

Unfortunately, obtaining an analytical expression for the posterior is typically not feasible. Firstly, the normalisation constant, $p(\bm{X})$, is unknown or determining it is nontrivial. Secondly, the likelihood, $p(\bm{X} | \bm{\theta})$, is intractable for most simulation models, particularly large-scale macroeconomic ABMs. Nevertheless, these limitations can be overcome to some extent. \citet{Grazzini_et_al_2017} provide a method for approximating $p(\bm{X} | \bm{\theta})$ for a particular value of $\bm{\theta}$, which then allows us to evaluate the right-hand side of
\begin{equation}
p(\bm{\theta} | \bm{X}) \propto p(\bm{X} | \bm{\theta})p(\bm{\theta}).
\end{equation}
The above may then be used along with Markov chain Monte Carlo (MCMC) methods, such as the Metropolis-Hastings algorithm, to sample the posterior. This is possible since most MCMC techniques only require that we are able to determine the value of a function proportional to the density function of interest rather than the density function itself. It should be apparent, however, that the overall estimation error will depend critically on the method used to approximate the likelihood.

\subsection{The Approach of \citet{Grazzini_et_al_2017}}

As previously stated, \citet{Grazzini_et_al_2017} provide a method to approximate the likelihood for simulation models, which we now discuss in more detail.

In essence, the approach is based on the assumption that, for all $t \geq \tilde{T}$, we reach a statistical equilibrium such that $\bm{x}_{t, i} ^{sim}(\bm{\theta})$ fluctuates around a stationary level, $\mathbb{E}[\bm{x}_{t, i} ^{sim}(\bm{\theta}) | t \geq \tilde{T}]$, which allows us to further assume that $\bm{x}_{\tilde{T}, i} ^{sim}(\bm{\theta}), \bm{x}_{\tilde{T} + 1} ^{sim}(\bm{\theta}), \dots, \bm{x}_{T, i} ^{sim}(\bm{\theta})$ constitutes a random sample from a given distribution\footnote{The samples need not all be drawn from a single Monte Carlo replication and may instead be drawn from the statistical equilibria reached by each replication in an ensemble generated using various random seeds. In practice, we simulate an ensemble of $R$ such Monte Carlo replications for each candidate set of $\bm{\theta}$ values and combine the samples from each replication into a single random sample.}. It is then possible to determine a density function that describes this distribution, which we denote by $\tilde{f}(\bm{x} | \bm{\theta})$, using kernel density estimation (KDE), finally allowing us to approximate the likelihood of the empirically-sampled data\footnote{Note that we have assumed, as in the case of the simulated data, that the empirically-sampled data fluctuates around a stationary level.} for a given value of $\bm{\theta}$ as follows:
\begin{equation}
p(\bm{X} | \bm{\theta}) = \prod_{t = 1} ^{T} \tilde{f}(\bm{x}_t | \bm{\theta}).
\end{equation}

It should be apparent that the above results in a simple strategy that is easy to apply in most contexts. It must be noted, however, that this is largely made possible through strong assumptions that seldom hold in practice. In more detail, notice that ordered time series (panel) data is essentially being treated as an i.i.d. random sample, implying that $\bm{x}_{t, i} ^{sim}(\bm{\theta}) \perp \bm{x}_{1, i} ^{sim}(\bm{\theta}), \dots, \bm{x}_{t - 1, i} ^{sim}(\bm{\theta})$ for all $t = 2, 3, \dots, T$. Unfortunately, such independence assumptions do hold for most simulation models, since $\bm{x}_{t, i} ^{sim}(\bm{\theta})$ is likely be dependent on at least some of the previously realised values, whether this dependence is explicit or mediated through latent variables. Additionally, such assumptions result in a likelihood function that makes no distinction between $\bm{\theta}$ values that result in identical unconditional distributions but differing temporal trends. Since many economic simulation models and particularly large-scale macroeconomic ABMs produce datasets that are characterised by seasonality or structural breaks, there is likely to be some impact on the quality of the resultant parameter estimates.

Nevertheless, \citet{Platt_2019} demonstrates that despite the above shortcomings, the method of \citet{Grazzini_et_al_2017} is able to provide reasonable parameter estimates in many contexts, while also outperforming several more sophisticated frequentist approaches. This warrants further investigation and naturally leads one to ask whether relaxing the required independence assumptions would allow for the construction of a superior Bayesian estimation method.

\subsection{Likelihood Approximation using Neural Networks}

We now begin our discussion of a relatively simple extension to the likelihood approximation procedure proposed by \citet{Grazzini_et_al_2017} that is capable of capturing some of the dependence of $\bm{x}_{t, i} ^{sim}(\bm{\theta})$ on past realised values. As a starting point, we assume that 
\begin{equation}
p \left(\bm{x}_{t, i} ^{sim} \big| \bm{x}_{1, i} ^{sim}, \dots, \bm{x}_{t - 1, i} ^{sim}: \bm{\theta}\right) =  p \left(\bm{x}_{t, i} ^{sim} \big| \bm{x}_{t - L, i} ^{sim}, \dots, \bm{x}_{t - 1, i} ^{sim}: \bm{\theta}\right) 
\end{equation}
for all $L < t \leq T$, implying that $\bm{x}_{t, i} ^{sim}(\bm{\theta})$ depends only on the past $L$ realised values. Our task, therefore, is the estimation of the above conditional densities, 
\begin{equation}
\tilde{f}\left(\bm{x}_{t - L, i} ^{sim}, \dots, \bm{x}_{t - 1, i} ^{sim}, \bm{x}_{t, i} ^{sim}, \bm{\phi} \right) \simeq p \left(\bm{x}_{t, i} ^{sim} \big| \bm{x}_{t - L, i} ^{sim}, \dots, \bm{x}_{t - 1, i} ^{sim}: \bm{\theta}\right),
\end{equation}
for all $L < t \leq T$, where $\bm{\phi} = \bm{\phi}(\bm{\theta})$ are parameters associated with the density estimation procedure.

In our context, we make use of a mixture density network (MDN), a neural network-based approach to conditional density estimation introduced by \citet{Bishop_1994}. The aforementioned scheme consists of two primary components\footnote{Note that these discussions are primarily illustrative and serve to briefly describe and motivate our approach. A detailed technical description of its implementation is provided in Appendix \ref{Method_Implementation}.}, a mixture of $K$ Gaussian random variables,
\begin{equation} \label{Network_Inference}
\tilde{f}\left(\bm{x}, \bm{y}, \bm{\phi} \right) = \sum_{k = 1} ^{K} \alpha_k \left(\bm{x}\right) \mathcal{N}\left(\bm{y} \big| \bm{\mu}_k\left(\bm{x}\right), \bm{\Sigma}_k\left(\bm{x}\right)\right),
\end{equation}
where we denote $\bm{x}_{t, i} ^{sim}$ by $\bm{y}$ and $\bm{x}_{t - L, i} ^{sim}, \dots, \bm{x}_{t - 1, i} ^{sim}$ by $\bm{x}$, and functions $\alpha_k$, $\bm{\mu}_k$ and $\bm{\Sigma}_k$ of $\bm{x}$ which determine the mixture parameters. Here, $\alpha_k$, $\bm{\mu}_k$ and $\bm{\Sigma}_k$ are the outputs of a feedforward neural network taking $\bm{x}$ as input and having weights and biases $\bm{\phi}(\bm{\theta})$, which are determined by training the network on an ensemble of $R$ Monte Carlo replications simulated by the candidate model for parameter set $\bm{\theta}$. Using the trained MDN, it is then possible to approximate the likelihood of the empirically-sampled data for a given value of $\bm{\theta}$ as follows:
\begin{equation}
p(\bm{X} | \bm{\theta}) = \prod_{t = 1} ^{T - L} \tilde{f}(\bm{x}_{t}, \dots, \bm{x}_{t + L - 1}, \bm{x}_{t + L}, \bm{\phi}).
\end{equation}

While alternative density estimation procedures could potentially have been employed, our consideration of MDNs is motivated primarily by their desirable properties. Specifically, MDNs are, in theory, capable of approximating fairly complex conditional distributions. This follows directly from the fact that mixtures of normal random variables are universal density approximators for sufficiently large $K$ \citep{Scott_2015} and the fact that neural networks are universal function approximators \citep{Hornik_1989}, provided they are sufficiently expressive. Therefore, provided that $K$ is sufficiently large and the constructed neural network sufficiently deep (and wide), the above methodology should result in accurate conditional density estimates.

\subsection{Method Comparison and Benchmarking} \label{Benchmarking}

Given that we have now described our proposed estimation methodology, we proceed to discuss our strategy for benchmarking it against the approach of \citet{Grazzini_et_al_2017}, where we follow a similar strategy to that employed in \citet{Platt_2019}.

We begin by letting $\bm{X}^{sim}(\bm{\theta}, T, i)$ be the output of a candidate model, $M$. Since empirically-observed data is nothing more than a single realisation of the true data-generating process, which may itself be viewed as a model with its own set of parameters, it follows that we may consider $\bm{X} = \bm{X}^{sim}(\bm{\theta}^{true}, T^{emp}, i^{*})$ as a proxy for real data to which $M$ may be calibrated.

In this case, we are essentially estimating a perfectly-specified model using data for which the true parameter values, $\bm{\theta}^{true}$, are known. It can be argued that a good estimation method would, in this idealised setting, be able to recover these true values to some extent and that methods which produce estimates closer to $\bm{\theta}^{true}$ would be considered superior. This leads us to define the following loss function
\begin{equation}
LS(\bm{\theta}^{true}, \hat{\bm{\theta}}) = || \bm{\theta}^{true} - \hat{\bm{\theta}} ||_{2},
\end{equation}
where $\hat{\bm{\theta}}$ is the parameter estimate (posterior mean) produced by a given Bayesian estimation method.

In practice, it is important that both $\hat{\bm{\theta}}$ and $\bm{\theta}^{true}$ are normalised to take values in the interval $[0, 1]$ before the loss function value is calculated. This is because even relatively small estimation errors associated with parameters that typically take on larger values will increase the loss function value substantially more than relatively large estimation errors associated with parameters that typically take on smaller values if no normalisation is performed. Therefore, for each free parameter, $\theta_j \in [a, b]$, we set
\begin{equation}
\hat{\theta}_{j} ^{[0, 1]} = \frac{\hat{\theta}_{j} - a}{b - a},
\end{equation}
with an analogous transformation being applied to $\theta ^{true} _{j}$.

The above allows us to develop a series of benchmarking exercises in which we compare the loss function values associated with our proposed method and that of \citet{Grazzini_et_al_2017} for a number of different models, free parameter sets, and $\bm{\theta}^{true}$ values\footnote{While the constructed loss function will act as our primary metric, we will also consider a number of other relevant criteria, such as the standard deviation of the obtained posteriors.}. In all of these comparative exercises, we aim to ensure that the overall conditions of the experiments are consistent throughout, regardless of the method used to approximate the likelihood. Therefore, in all cases, we set the length of the proxy for real data to be $T_{emp} = 1000$, the number of Monte Carlo replications in the simulated ensembles to be $R = 100$, the length of each series in the simulated ensembles to be $T_{sim} = 1000$, and the priors for all free parameters to be uniform over the explored parameter ranges. Additionally, we have also used the same lag length, $L = 3$, for all estimation attempts involving our neural network-based method. While seemingly arbitrary, this choice has very clear motivations that are discussed in detail in Section \ref{Lag_Robustness_Demo}.

Finally, the MCMC algorithm used to sample the posterior and its associated hyperparameters remain unchanged in all experiments. Rather than using a standard random walk Metropolis-Hastings algorithm, we have instead employed the adaptive scheme proposed by \citet{Griffin_Walker_2013}, which allows for more effective initialisation, faster convergence, and better handling of multimodal posteriors\footnote{A complete description of the procedure is presented in Appendix \ref{MCMC_Alg}.}.

\section{Candidate Models}

With our estimation and benchmarking strategies now described, we introduce the candidate models that we attempt to estimate. Their selection is primarily justified by their ubiquity; each has appeared in a number of calibration and estimation studies\footnote{For example, the \citet{Brock_Hommes_1998} model is considered by \citet{Recchioni_et_al_2015}, \citet{Lamperti_et_al_2018}, and \citet{Kukacka_Barunik_2017} and the \citet{Franke_Westerhoff_2012} model is considered by \citet{Franke_Westerhoff_2012} and \citet{Lux_2018}.}, leading them to become standard test cases in the field. While computationally-inexpensive to simulate, most are capable of producing nuanced dynamics and thus still prove to be a reasonable challenge for most contemporary estimation approaches. Since our focus here is the benchmarking of the proposed estimation procedure as opposed to estimating the candidate models using empirical data, our discussion will be relatively brief. In empirical investigations, however, it would be necessary to provide some justification that the chosen models were reasonable representations of the considered systems.

\subsection{\citet{Brock_Hommes_1998} Model}

The first model we introduce, and by far the most popular in the existing literature, is the \citet{Brock_Hommes_1998} model, an early example of a class of simulation models that attempt to model the trading of assets on an artificial stock market by simulating the interactions of heterogenous traders that follow various trading strategies. 

We focus on a particular version of the model that can be expressed as a system of coupled equations\footnote{The interested reader should refer to \citet{Brock_Hommes_1998} for a detailed discussion of the model's underlying assumptions and the derivation of the above system of equations.},
\begin{align}
y_{t + 1} &= \frac{1}{1 + r} \sum_{h = 1} ^{H} n_{h, t + 1} (g_h y_t + b_h) + \epsilon_{t + 1} \text{, } \epsilon_t \sim \mathcal{N}(0, \sigma^2), \\
n_{h, t + 1} &= \frac{\exp(\beta U_{h, t})}{\sum_{h = 1} ^{H}\exp(\beta U_{h, t})}, \\
U_{h, t} &= (y_t - R y_{t - 1})(g_h y_{t - 2} + b_h - R y_{t - 1}),
\end{align}
where $y_{t}$ is the asset price at time $t$ (in deviations from the fundamental value $p_{t} ^{*}$), $n_{h, t}$ is the fraction of trader agents following strategy $h \in \left\{1, 2, \dots, H \right\}$ at time $t$, and $R = 1 + r$.

Each strategy, $h$, has an associated trend following component, $g_h$, and bias, $b_h$, both of which are real-valued parameters. The model also includes positive-valued parameters that affect all trader agents, regardless of the strategy they are currently employing, specifically $\beta$, which controls the rate at which agents switch between various strategies, and the prevailing market interest rate, $r$.

Finally, assuming an i.i.d. dividend process, the fundamental value $p_{t} ^{*} = p^{*}$ is constant, allowing us to obtain the asset price at time $t$,
\begin{equation}
p_t = y_t + p^{*}.
\end{equation}

\subsection{Random Walks with Structural Breaks}

The second model we consider is a random walk capable of replicating simple structural breaks, defined according to
\begin{equation}
x_{t + 1}= x_{t} + d_{t + 1} + \epsilon_{t + 1} \text{, } \epsilon_t \sim \mathcal{N}(0, \sigma_t ^2),
\end{equation}
where
\begin{equation}
d_t, \sigma_t = 
\begin{cases}
d_1, \sigma_1 & t \leq \tau \\
d_2, \sigma_2 & t > \tau.
\end{cases}
\end{equation}

Unlike the \citet{Brock_Hommes_1998} model, the above is not a representation of a real-world system, but rather an artificially-constructed test example designed to challenge estimation methodologies\footnote{This particular instantiation of the model was first used by \citet{Lamperti_2017} to test an information-theoretic criterion called the GSL-div.}. Its inclusion is justified on the grounds that, as previously discussed, many large-scale ABMs produce dynamics that are characterised by structural breaks and the fact that it allows us to compare our approach against that of \citet{Grazzini_et_al_2017} in cases where the considered data demonstrates clear temporal changes in the prevailing dynamics.

\subsection{\citet{Franke_Westerhoff_2012} Model} \label{Franke_Westerhoff_Model}

The final model we discuss shares a number of conceptual similarities with the previously described \citet{Brock_Hommes_1998} model, being a heterogeneous agent model that simulates the interactions of traders following a number of trading strategies. It is, however, different in a number of key areas, particularly in how the probability of an agent switching from one strategy to another is determined and in its incorporation of only two trader types, chartists and fundamentalists.

As in the case of the \citet{Brock_Hommes_1998} model, the core elements of the model can be expressed as a system of coupled equations
\begin{align} \label{Franke_Westerhoff_Main_Start}
p_t &= p_{t - 1} + \mu \left(n_{t - 1} ^{f} d_{t - 1} ^{f} + n_{t - 1} ^{c} d_{t - 1} ^{c} \right),  \\
d_{t} ^{f} &= \phi (p^{*} - p_t) + \epsilon_{t} ^{f} \text{, } \epsilon_{t} ^{f} \sim \mathcal{N}(0, \sigma_{f} ^{2}),\\
d_{t} ^{c} &= \chi (p_{t} - p_{t - 1}) + \epsilon_{t} ^{c} \text{, } \epsilon_{t} ^{c} \sim \mathcal{N}(0, \sigma_{c} ^{2}),\\
n_{t} ^{f} &= \frac{1}{1 + \exp(-\beta a_{t - 1})},\\
n_{t} ^{c} &= 1 - n_{t} ^{f}, \label{Franke_Westerhoff_Main_End}
\end{align}
where $p_t$ is the log asset price at time $t$, $p^{*}$ is the log of the (constant) fundamental value, $n_{t} ^{f}$ and $n_{t} ^{c}$ are the market fractions of fundamentalists and chartists respectively at time $t$, $d_{t} ^{f}$ and $d_{t} ^{c}$ are the corresponding average demands, and the remaining symbols all correspond to positive-valued parameters.

At this point, it is worth pointing out that \citet{Franke_Westerhoff_2012} do not introduce a single model, but rather a family of related formulations built on the same foundation (Eqns. \ref{Franke_Westerhoff_Main_Start}-\ref{Franke_Westerhoff_Main_End}). These models differ in how they define $a_{t}$, the attractiveness of fundamentalism relative to chartism at the end of period $t$, and incorporate a number of different mechanisms, including wealth, herding and price misalignment. This makes the consideration of multiple versions of the model worthwhile and we thus consider two of the proposed versions\footnote{$\alpha_n$, $\alpha_w$, and $\alpha_p$ are strictly positive while $\alpha_0$ may take on any real value.}:
\begin{equation}
a_t = \alpha_n (n_{t} ^ {f} - n_{t} ^{c}) + \alpha_0 + \alpha_p (p_t - p^{*}) ^{2},
\end{equation}
referred to as herding, predisposition and misalignment (HPM), and
\begin{align}
g_{t} ^{s} &= \left[\exp(p_t) - \exp(p_{t - 1})\right] d_{t - 2} ^{s} \text{, } s = \{f, c\}, \label{Wealth_1} \\
w_{t} ^{s} &= \eta w_{t - 1} ^{s} + (1 - \eta) g_{t} ^{s}, \\
a_t &= \alpha_w (w_{t} ^ {f} - w_{t} ^{c}) + \alpha_0, \label{Wealth_2}
\end{align}
referred to as wealth and predisposition (WP).

As a final remark, we consider $r_t = p_t - p_{t - 1}$, the log return process, rather than $p_t$ in our estimation attempts. 

\section{Results and Discussion}

\subsection{\citet{Brock_Hommes_1998} Model}

We now proceed with the presentation of the results of our comparative experiments, beginning with the \citet{Brock_Hommes_1998} model\footnote{From this point onwards, we use KDE to refer to the method of \citet{Grazzini_et_al_2017} and MDN to refer to our proposed method in all tables and figures.}. 

In these experiments, we consider a market with $H = 4$ trading strategies and focus on estimating $g_2$, $b_2$, $g_3$, and $b_3$, the trend following and bias components for two of these strategies. For the first free parameter set, we consider $g_2, b_2 \in [-1, 0]$ and $g_3, b_3 \in [0, 1]$, corresponding to a contrarian strategy with a negative bias and a trend following strategy with a positive bias respectively. For the second free parameter set, we instead consider $g_2, b_2, g_3 \in [0, 1]$ and $b_3 \in [-1, 0]$, corresponding to trend following strategies with positive and negatives biases respectively.

\begin{figure}[H]

\centering

\makebox[\linewidth][c]{%
\begin{subfigure}{0.55\linewidth}
	\centering
	\includegraphics[width=1\linewidth]{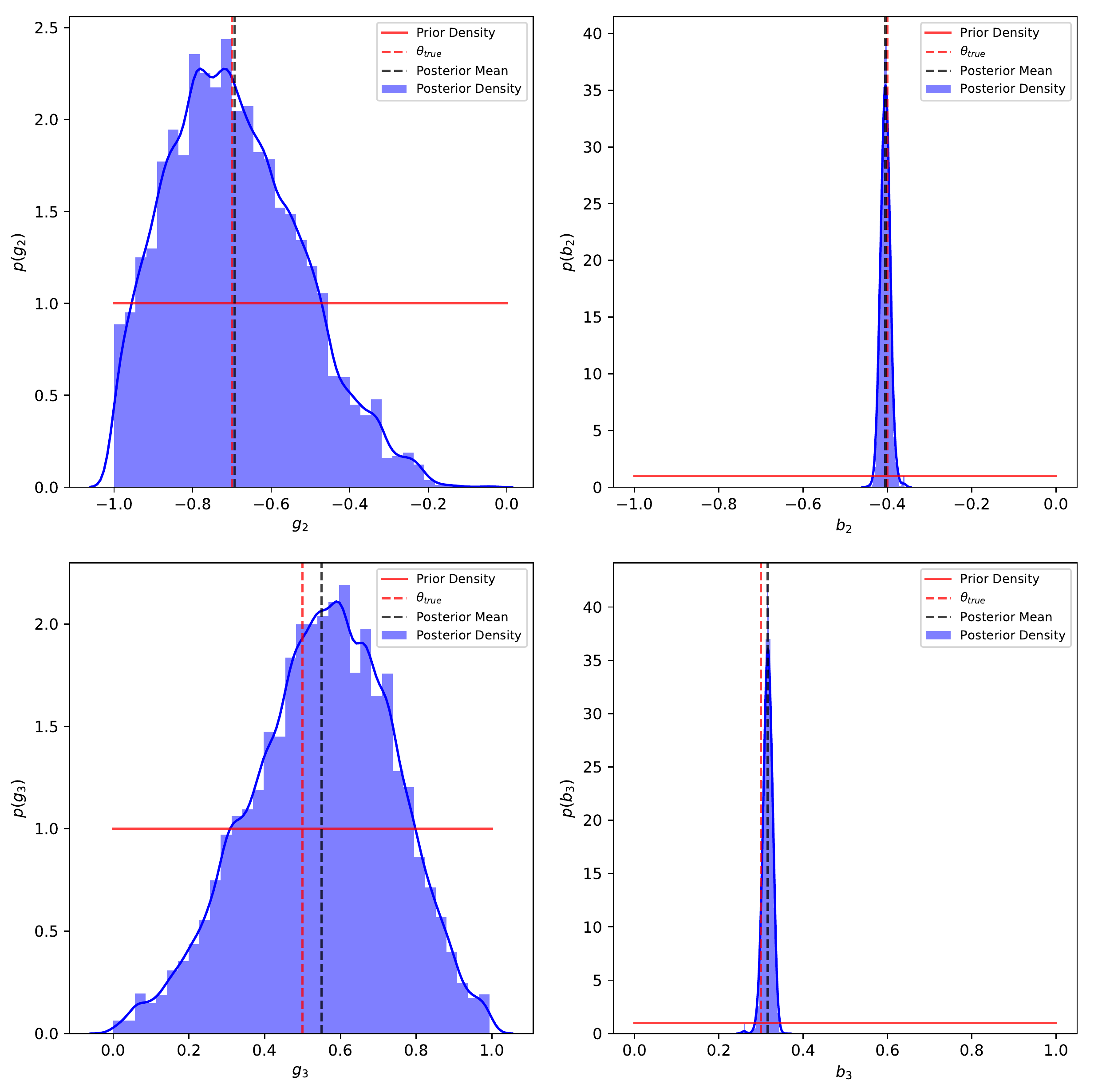}
	\caption{MDN}
\end{subfigure}
\begin{subfigure}{0.55\linewidth}
	\centering
	\includegraphics[width=1\linewidth]{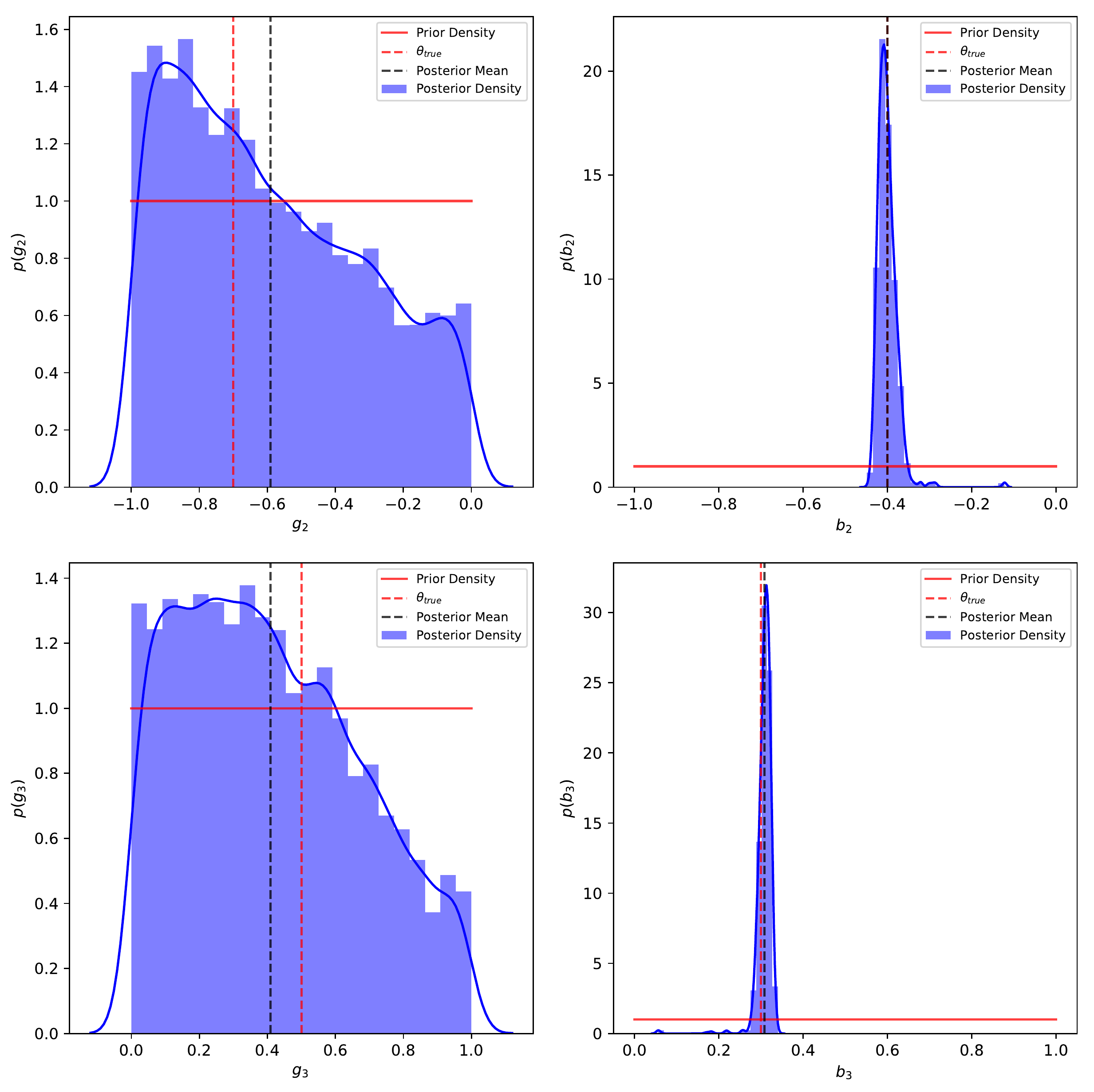}
	\caption{KDE}
\end{subfigure}
}

\caption{Marginal posterior distributions for free parameter set $1$ of the \citet{Brock_Hommes_1998} model.} \label{BH_1_Posterior}

\end{figure}

Referring to Figure \ref{BH_1_Posterior}, we observe that, for the first free parameter set, there is a pronounced difference in performance between our proposed methodology and that of \citet{Grazzini_et_al_2017}. While both approaches perform similarly when estimating the bias components, our proposed procedure results in marginal posteriors for $g_2$ and $g_3$ that not only have means noticeably closer to the true parameter values, but are also significantly narrower and more peaked, with their density concentrated in a smaller region of the parameter space. This can be seen as indicative of reduced estimation uncertainty.

Table \ref{BH_Summary} elaborates on these findings and reveals that similar behaviours also emerge in the case of the second free parameter set. Specifically, we find that the posterior means ($\bm{\mu}_{posterior}$) for both methods result in more or less equivalent estimates for $b_2$ and $b_3$, while the posterior mean for our proposed method appears to result in noticeably superior estimates for $g_2$ and $g_3$ in both cases, ultimately leading to lower loss function values. We also observe that our approach results in reduced posterior standard deviations ($\bm{\sigma}_{posterior}$) consistently for all free parameters, in line with our observation of reduced estimation uncertainty in Figure \ref{BH_1_Posterior}.

\begin{table}[H]

\caption{Estimation Result Summary for the \citet{Brock_Hommes_1998} Model} \label{BH_Summary}

\begin{tabularx}{\linewidth}{lXXXX}

\hline
\midrule
 & $g_2$ & $b_2$ & $g_3$ & $b_3$ \\
 
\hline
\midrule
\textit{Param Set} $1$ & & & & \\
$\bm{\theta}_{true}$ & $-0.7$ & $-0.4$ &  $0.5$ & $0.3$ \\

\midrule
\textit{MDN} \\
$\bm{\mu}_{posterior}$ & $-0.6931$ & $-0.4048$ & $0.5505$ & $0.3160$ \\
$\bm{\sigma}_{posterior}$ & $0.1681$ & $0.0105$ & $0.1864$ & $0.0103$ \\
$\bm{\sigma}_{sampling}$ & $0.0051$ & $0.0002$ & $0.0055$ & $0.0003$ \\
$LS$ & \multicolumn{4}{c}{\hspace{-1.35cm}$\bm{0.0536}$} \\

\midrule
\textit{KDE} \\
$\bm{\mu}_{posterior}$ & $-0.5910$ &  $-0.4004$ & $0.4092$ & $0.3083$ \\
$\bm{\sigma}_{posterior}$ & $0.2787$ & $0.0254$ & $0.2603$ & $0.0197$ \\
$\bm{\sigma}_{sampling}$ & $0.0089$ & $0.0012$ & $0.0130$ & $0.0011$ \\
$LS$ & \multicolumn{4}{c}{\hspace{-1.35cm}$\bm{0.1421}$} \\

\hline
\midrule
\textit{Param Set} $2$ & & & & \\
$\bm{\theta}_{true}$ & $0.6$ & $0.2$ & $0.7$ & $-0.2$ \\

\midrule
\textit{MDN} \\
$\bm{\mu}_{posterior}$ & $0.6021$ & $0.2401$ & $0.7493$ & $-0.2304$ \\
$\bm{\sigma}_{posterior}$ & $0.1804$ & $0.0149$ & $0.1662$ & $0.0147$ \\
$\bm{\sigma}_{sampling}$ & $0.0116$ & $0.0004$ & $0.0090$ & $0.0004$ \\
$LS$ & \multicolumn{4}{c}{\hspace{-1.35cm}$\bm{0.0705}$} \\

\midrule
\textit{KDE} \\
$\bm{\mu}_{posterior}$ & $0.4658$ & $0.2410$ & $0.6461$ & $-0.2330$ \\
$\bm{\sigma}_{posterior}$ & $0.2803$ & $0.0677$ & $0.2571$ & $0.0666$ \\
$\bm{\sigma}_{sampling}$ & $0.01693$ &  $0.0067$ & $0.0145$ & $0.0067$ \\
$LS$ & \multicolumn{4}{c}{\hspace{-1.35cm}$\bm{0.1539}$} \\
\midrule
\hline

\end{tabularx}

\caption*{$g_0 = b_0 = b_4 = 0$, $g_4 = 1.01$, $r = 0.01$, $\beta = 10$, and $\sigma = 0.04$ for both free parameter sets.}

\end{table}

In Appendix \ref{MCMC_Alg}, where we describe the method used to sample the posteriors, we indicate that we run the procedure multiple times with different initial conditions and combine the obtained samples into a single, larger sample from which we estimate $\bm{\mu}_{posterior}$ and $\bm{\sigma}_{posterior}$. We can, however, estimate the posterior mean for each of these runs individually and determine the standard deviation of $\bm{\mu}_{posterior}$ across the instantiations of the algorithm, which we call $\bm{\sigma}_{sample}$. As shown in Table \ref{BH_Summary}, this standard deviation is generally very small for both methods, suggesting that the posterior mean estimates are generally robust\footnote{This is true for all free parameter sets and models considered in this investigation.}.

\subsection{Random Walks with Structural Breaks}

Moving on from the \citet{Brock_Hommes_1998} model, we now discuss the estimation of a random walk incorporating a structural break. In these experiments, we consider a fixed structural break location, $\tau = 700$\footnote{This induces a degree of asymmetry in the data and results in a more challenging and realistic estimation problem than $\tau = 500$.}, and determine the extent to which both methods are capable of estimating the pre- and post-break drift, $d_1, d_2 \in [0, 1]$, and volatility, $\sigma_1, \sigma_2 \in [0, 10]$, for differing underlying changes in the dynamics. While the loss function described in Section \ref{Benchmarking} will still be used as our primary metric, we note that since the considered free parameters directly define the dynamics that characterise the different regimes of the data, it would also be worthwhile to assess the extent to which the competing approaches are able to correctly identify the relationships between the parameters and hence the shift in the pre- and post-break dynamics ($\Delta_d$ and $\Delta_{\sigma}$).

\begin{table}[H]

\caption{Estimation Result Summary for the Random Walk Model (Increasing Volatility)} \label{RW_Volatility}

\begin{tabularx}{\linewidth}{lXXX}

\hline
\midrule
 & $\sigma_1$ & $\sigma_2$ & $\Delta_{\sigma}$ \\
 
\hline
\midrule
\textit{Param Set} $1$ & & & \\
$\bm{\theta}_{true}$ & $1$ & $2$ & $1$ \\

\midrule
\textit{MDN} \\
$\bm{\mu}_{posterior}$ & $1.0585$ & $1.9957$ & $0.9372$ \\
$\bm{\sigma}_{posterior}$ & $0.8153$ & $0.6517$ & $-$ \\
$\bm{\sigma}_{sampling}$ & $0.0137$ & $0.0629$ & $-$ \\
$LS$ & & $\bm{0.0059}$ & \\

\midrule
\textit{KDE} \\
$\bm{\mu}_{posterior}$ & $0.9966$ & $1.9084$ & $0.9118$ \\
$\bm{\sigma}_{posterior}$ & $0.4113$ & $0.2719$ & $-$ \\
$\bm{\sigma}_{sampling}$ & $0.0430$ & $0.0197$ & $-$ \\
$LS$ & & $\bm{0.0092}$ & \\

\hline
\midrule
\textit{Param Set} $2$ & & & \\
$\bm{\theta}_{true}$ & $1$ & $2$ & $1$ \\

\midrule
\textit{MDN} \\
$\bm{\mu}_{posterior}$ & $1.0205$ & $1.9598$ & $0.9393$ \\
$\bm{\sigma}_{posterior}$ & $0.5660$ & $0.4605$ & $-$ \\
$\bm{\sigma}_{sampling}$ & $0.0216$ & $0.0478$ & $-$ \\
$LS$ & & $\bm{0.0045}$ & \\

\midrule
\textit{KDE} \\
$\bm{\mu}_{posterior}$ & $0.9790$ & $1.8930$ & $0.9144$ \\
$\bm{\sigma}_{posterior}$ & $0.0923$ & $0.2141$ & $-$ \\
$\bm{\sigma}_{sampling}$ & $0.0046$ & $0.0169$ & $-$ \\
$LS$ & & $\bm{0.0109}$ & \\
\midrule
\hline

\end{tabularx}

\caption*{$d_1 = 0.4$ and $d_2 = 0.5$ for free parameter set $1$ and $d_1 = 0.1$ and $d_2 = 0.2$ for free parameter set $2$.}

\end{table}

Before proceeding, however, there are a number of nuances that should be highlighted. Being a random walk, the model clearly produces non-stationary time series and therefore violates a key assumption of the method of \citet{Grazzini_et_al_2017}. For this reason, it is necessary to consider the series of first differences, $x_t - x_{t - 1}$, rather than $x_t$ itself. While our approach does not make stationarity assumptions, we have none the less considered the series of first differences when applying both methods to make the comparison as fair as possible. It should also be noted that we have assumed the location of the structural break to be unknown or difficult to determine a-priori (as is the case in most practical problems), meaning that we apply both estimation approaches to the full time series data to estimate both the pre- and post-break parameters simultaneously. If, however, the location of the structural break was known, it would be possible to estimate the relevant parameters separately using appropriate subsets of the data, a less challenging undertaking that we do not consider here.

Now, referring to Table \ref{RW_Volatility}, we see that both our proposed estimation methodology and that of \citet{Grazzini_et_al_2017} perform similarly well when attempting to estimate the pre- and post-break volatility, with both producing reasonable estimates for the free parameters and both being able to identity the correct shift in the dynamics. Referring to Tables \ref{RW_Drift_Increase} and \ref{RW_Drift_Decrease}, however, we see that more pronounced differences emerge when attempting to estimate the pre- and post-break drift. While this is clearly evident from the fact that the loss function values associated with our proposed methodology are noticeably lower in all cases, a more detailed analysis reveals further distinctions worth mentioning. Table \ref{RW_Drift_Increase}, which presents the results for cases involving an increasing drift, reveals that our proposed methodology has correctly identified an increasing trend in both cases and has also correctly identified that the increase in drift for parameter set $4$ is three times that of parameter set $3$. In contrast to this, the method of \citet{Grazzini_et_al_2017} incorrectly suggests a decreasing trend in both cases. Table \ref{RW_Drift_Decrease}, which presents the results for cases involving a decreasing drift, similarly shows that our proposed methodology delivers superior performance when attempting to identify the change in drift. 

\begin{table}[H]

\caption{Estimation Result Summary for the Random Walk Model (Increasing Drift)} \label{RW_Drift_Increase}

\begin{tabularx}{\linewidth}{lXXX}

\hline
\midrule
 & $d_1$ & $d_2$ & $\Delta_{d}$ \\
 
\hline
\midrule
\textit{Param Set} $3$ & & & \\
$\bm{\theta}_{true}$ & $0.4$ & $0.5$ & $0.1$ \\

\midrule
\textit{MDN} \\
$\bm{\mu}_{posterior}$ & $0.4867$ & $0.5465$ & $0.0598$ \\
$\bm{\sigma}_{posterior}$ & $0.0536$ & $0.1139$ & $-$ \\
$\bm{\sigma}_{sampling}$ & $0.0056$ & $0.0038$ & $-$ \\
$LS$ & & $\bm{0.0984}$ & \\

\midrule
\textit{KDE} \\
$\bm{\mu}_{posterior}$ & $0.5204$ & $0.3258$ & $-0.1945$ \\
$\bm{\sigma}_{posterior}$ & $0.0578$ & $0.1463$ & $-$ \\
$\bm{\sigma}_{sampling}$ & $0.0032$ & $0.0050$ & $-$ \\
$LS$ & & $\bm{0.2117}$ & \\

\hline
\midrule
\textit{Param Set} $4$ & & & \\
$\bm{\theta}_{true}$ & $0.4$ & $0.7$ & $0.3$ \\

\midrule
\textit{MDN} \\
$\bm{\mu}_{posterior}$ & $0.5054$ & $0.6876$ & $0.1823$ \\
$\bm{\sigma}_{posterior}$ & $0.0434$ & $0.1131$ & $-$ \\
$\bm{\sigma}_{sampling}$ & $0.0024$ & $0.0036$ & $-$ \\
$LS$ & & $\bm{0.1061}$ & \\

\midrule
\textit{KDE} \\
$\bm{\mu}_{posterior}$ & $0.5308$ & $0.5033$ & $-0.0275$ \\
$\bm{\sigma}_{posterior}$ & $0.0561$ & $0.1457$ & $-$ \\
$\bm{\sigma}_{sampling}$ & $0.0025$ & $0.0041$ & $-$ \\
$LS$ & & $\bm{0.2362}$ & \\
\midrule
\hline

\end{tabularx}

\caption*{$\sigma_1 = 1$ and $\sigma_2 = 2$ for both free parameter sets.}

\end{table}

This change in the relative performances of each method when estimating the drift rather than the volatility is a direct consequence of the relationship between the deterministic and stochastic components of the model. For the selected parameter ranges, the random fluctuations, $\epsilon_t$, dominate the evolution of the model, with the drift producing a more subtle effect, particularly after the structural break occurs. For this reason, correctly estimating the pre- and post-break volatility is a far less challenging task than estimating the pre- and post-break drift. Therefore, while both methods perform well when estimating parameters associated with dominant effects like volatility, our method's incorporation of dependence on previously observed values seems to be important when estimating parameters related to more nuanced and less dominant aspects of a model.

\begin{table}[H]

\caption{Estimation Result Summary for the Random Walk Model (Decreasing Drift)} \label{RW_Drift_Decrease}

\begin{tabularx}{\linewidth}{lXXX}

\hline
\midrule
 & $d_1$ & $d_2$ & $\Delta_{d}$ \\
 
\hline
\midrule
\textit{Param Set} $5$ & & & \\
$\bm{\theta}_{true}$ & $0.5$ & $0.4$ & $-0.1$ \\

\midrule
\textit{MDN} \\
$\bm{\mu}_{posterior}$ & $0.5691$ & $0.4743$ & $-0.0949$ \\
$\bm{\sigma}_{posterior}$ & $0.0485$ & $0.1348$ & $-$ \\
$\bm{\sigma}_{sampling}$ & $0.0031$ & $0.0039$ & $-$ \\
$LS$ & & $\bm{0.1015}$ & \\

\midrule
\textit{KDE} \\
$\bm{\mu}_{posterior}$ & $0.6015$ & $0.2611$ & $-0.3404$ \\
$\bm{\sigma}_{posterior}$ & $0.0573$ & $0.1396$ & $-$ \\
$\bm{\sigma}_{sampling}$ & $0.0039$ & $0.0032$ & $-$ \\
$LS$ & & $\bm{0.1720}$ & \\

\hline
\midrule
\textit{Param Set} $6$ & & & \\
$\bm{\theta}_{true}$ & $0.7$ & $0.4$ & $-0.3$ \\

\midrule
\textit{MDN} \\
$\bm{\mu}_{posterior}$ & $0.7585$ & $0.4400$ & $-0.3185$ \\
$\bm{\sigma}_{posterior}$ & $0.0532$ & $0.1526$ & $-$ \\
$\bm{\sigma}_{sampling}$ & $0.0033$ & $0.0029$ & $-$ \\
$LS$ & & $\bm{0.0709}$ & \\

\midrule
\textit{KDE} \\
$\bm{\mu}_{posterior}$ & $0.7838$ & $0.2934$ & $-0.4904$ \\
$\bm{\sigma}_{posterior}$ & $0.0564$ & $0.1469$ & $-$ \\
$\bm{\sigma}_{sampling}$ & $0.0027$ & $0.0030$ & $-$ \\
$LS$ & & $\bm{0.1356}$ & \\
\midrule
\hline

\end{tabularx}

\caption*{$\sigma_1 = 1$ and $\sigma_2 = 2$ for both free parameter sets.}

\end{table}

\subsection{\citet{Franke_Westerhoff_2012} Model}

As stated in Section \ref{Franke_Westerhoff_Model}, the final model we consider has a number of alternate configurations differing in how the attractiveness of fundamentalism relative to chartism, $a_t$, is determined during each period. For this reason, we consider two of these configurations, HPM and WP, and focus on estimating the parameters associated with the rules governing $a_t$: $\alpha_n \in [0, 2]$, $\alpha_0 \in [-1, 1]$, $\alpha_p \in [0, 20]$, $\alpha_w \in [0, 15000]$, and $\eta \in [0, 1]$, while also estimating the standard deviation of the noise term appearing in the chartist demand equation, $\sigma_c \in [0, 5]$\footnote{We originally attempted to estimate $\sigma_f$ as well, but found this to exhibit a degree of collinearity with $\sigma_c$.}.

Referring to Table \ref{FW_Summary}, we see that our proposed estimation methodology appears slightly more effective than that of \citet{Grazzini_et_al_2017} for the HPM parameter set, producing superior estimates for all but one of the considered free parameters and resulting in a lower loss function value. Nevertheless, the estimates do not differ substantially when comparing the methods. Despite this, we see, in what is a seemingly analogous trend to what was observed in the random walk experiments, that the differences in performance are more pronounced for the WP parameter set. In particular, we see a substantial difference in the loss function values associated with each method, brought about by differences in the quality of estimates produced for $\eta$.

\begin{table}[H]

\caption{Estimation Result Summary for the \citet{Franke_Westerhoff_2012} Model} \label{FW_Summary}

\begin{tabularx}{\linewidth}{lXXXX}

\hline
\midrule
 & $\alpha_0$ & $\alpha_n$ & $\alpha_p$ & $\sigma_c$ \\
 
\hline
\midrule
\textit{Param Set HPM} & & & & \\
$\bm{\theta}_{true}$ & $-0.327$ & $1.79$ & $18.43$ & $2.087$ \\

\midrule
\textit{MDN} \\
$\bm{\mu}_{posterior}$ & $-0.1749$ & $1.8987$ & $17.1821$ & $2.3113$ \\
$\bm{\sigma}_{posterior}$ & $0.1297$ & $0.1697$ & $2.2932$ & $0.3548$ \\
$\bm{\sigma}_{sampling}$ & $0.0036$ & $0.0232$ & $0.0410$ &  $0.0130$ \\
$LS$ & \multicolumn{4}{c}{\hspace{-1.35cm}$\bm{0.1210}$} \\

\midrule
\textit{KDE} \\
$\bm{\mu}_{posterior}$ & $-0.1287$ & $1.7968$ & $16.2177$ & $2.3134$ \\
$\bm{\sigma}_{posterior}$ & $0.1667$ & $0.2880$ & $3.1280$ & $0.5547$ \\
$\bm{\sigma}_{sampling}$ & $0.0139$ & $0.0105$ & $0.2356$ & $0.05105$ \\
$LS$ & \multicolumn{4}{c}{\hspace{-1.35cm}$\bm{0.15534}$} \\

\hline
\midrule
 & $\alpha_w$ & $\eta$ & $\sigma_c$ & \\
 
\hline
\midrule
\textit{Param Set WP} & & & & \\
$\bm{\theta}_{true}$ & $2668$ & $0.987$ & $1.726$ & \\

\midrule
\textit{MDN} \\
$\bm{\mu}_{posterior}$ & $1993.1311$ & $0.9078$ & $1.6991$ & \\
$\bm{\sigma}_{posterior}$ & $2195.8553$ & $0.0799$ & $0.4335$ & \\
$\bm{\sigma}_{sampling}$ & $184.4589$ & $0.0043$ & $0.0364$ & \\
$LS$ & \multicolumn{4}{c}{\hspace{-1.35cm}$\bm{0.0912}$} \\

\midrule
\textit{KDE} \\
$\bm{\mu}_{posterior}$ & $2437.1697$ & $0.6263$ & $1.4567$ & \\
$\bm{\sigma}_{posterior}$ & $2831.5574$ & $0.2846$ & $0.3403$ & \\
$\bm{\sigma}_{sampling}$ & $458.0461$ & $0.0257$ & $0.0296$ & \\
$LS$ & \multicolumn{4}{c}{\hspace{-1.35cm}$\bm{0.3650}$} \\
\midrule
\hline

\end{tabularx}

\caption*{$\mu = 0.01$, $\beta = 1$, $\phi = 0.12$, $\chi = 1.5$, and $\sigma_f = 0.758$ for the HPM parameter set and $\mu = 0.01$, $\beta = 1$, $\phi = 1$, $\chi = 0.9$, $\alpha_0 = 2.1$, and $\sigma_f = 0.752$ for the WP parameter set, as suggested by \citet{Franke_Westerhoff_2012}.}

\end{table}

\begin{figure}[H]

\centering
\begin{subfigure}{1\linewidth}
	\centering
	\includegraphics[width=1\linewidth]{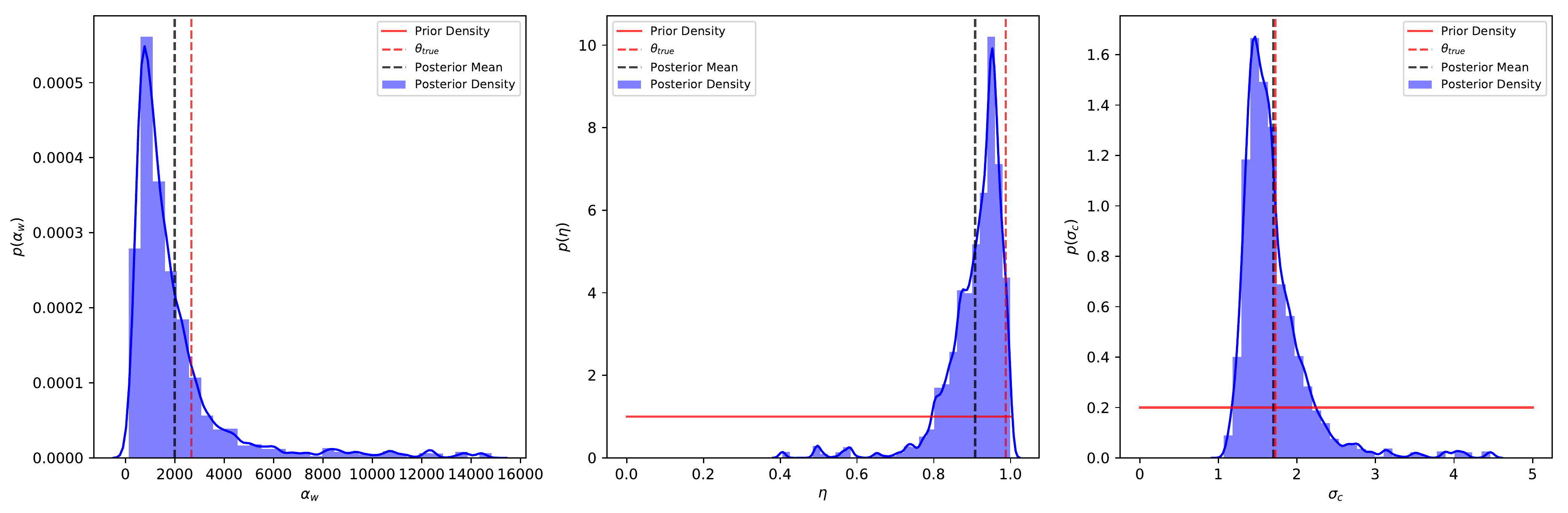}
	\caption{MDN}
\end{subfigure}
\begin{subfigure}{1\linewidth}
	\centering
	\includegraphics[width=1\linewidth]{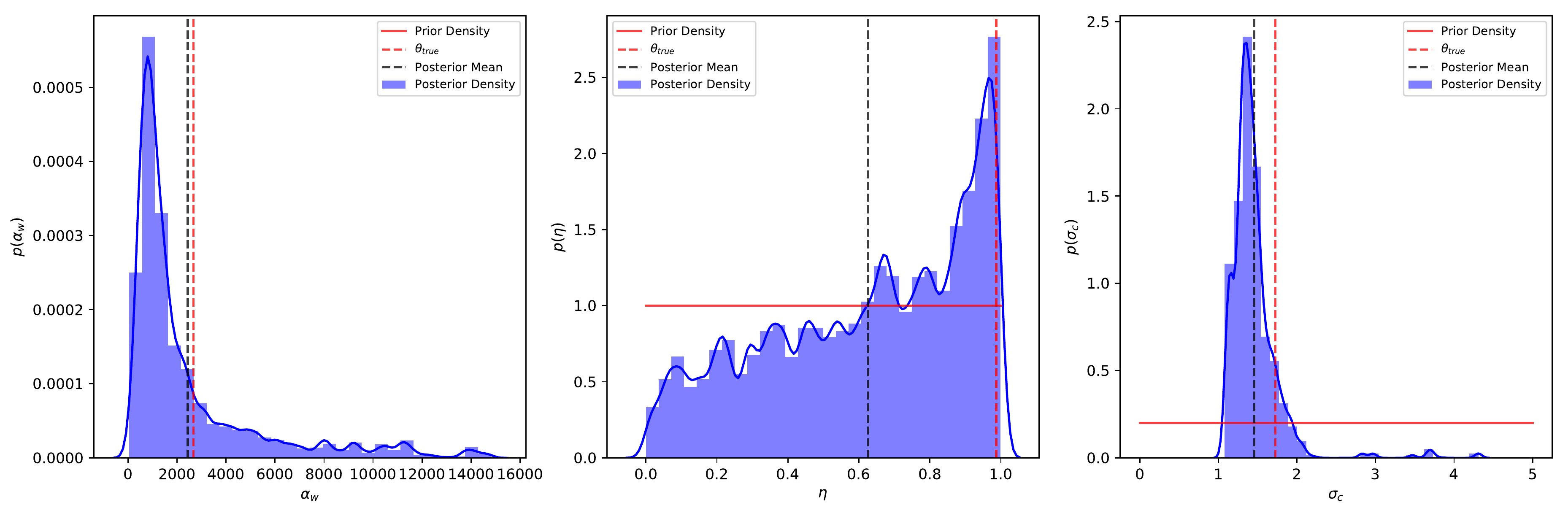}
	\caption{KDE}
\end{subfigure}

\caption{Marginal posterior distributions for the WP parameter set of the \citet{Franke_Westerhoff_2012} model.} \label{FW_WP}

\end{figure}

As illustrated in Figure \ref{FW_WP}, the method of \citet{Grazzini_et_al_2017} produces a wide posterior for $\eta$ that is dispersed across the entirety of the explored parameter range, which results in a relatively poor estimate. In contrast to this, we see that the proposed methodology fares better, producing a far narrower posterior and a significantly more accurate estimate. While it is nontrivial to identify any definitive causes for the observed behaviours due to the nonlinear nature of heterogeneous agent models, it is worth pointing out that the inclusion of wealth dynamics in the WP version of the model introduces a dependence of $a_t$ on the previous return via Eqns. \ref{Wealth_1}-\ref{Wealth_2}, which may in turn increase the strength of the relationship between the current and previously observed values in the log return time series.

As a final remark, notice that for the vast majority of the free parameters considered, the proposed methodology also results in lower posterior standard deviations, as was the case for the \citet{Brock_Hommes_1998} model.

\subsection{Overall Summary}

In the preceding subsections, we have focused primarily on analysing the results on a case-by-case basis. Here, however, we provide a summative comparison across all of the considered models. This is achieved though the consideration of a number of key performance metrics, presented in Table \ref{Overall_Summary}, which compare the approaches at both a global and individual parameter level. 

The first of the aforementioned metrics, and the most important, $LS_{mdn} < LS_{kde}$, indicates how often the proposed methodology results in lower loss function values, and hence measures its relative ability to recover the true parameter set. We observe that in all cases considered, our methodology results in lower loss function values, which can be seen as indicative of dominance at the global level.

\begin{table}[H]

\setlength\extrarowheight{3pt}

\caption{Estimation Result Summary Across All Models} \label{Overall_Summary}

\begin{tabularx}{\linewidth}{XX}

\hline
\midrule
Outcome & Percentage of Cases \\
\midrule
\hline
$LS_{mdn} < LS_{kde}$ & $100$ \\
$|\mu_{mdn} ^{i} - \theta_{true} ^{i}| < |\mu_{kde} ^{i} - \theta_{true} ^{i}|$ & $81.48$ \\
$\sigma_{mdn} ^{i} < \sigma_{kde} ^{i}$ & $77.78$ \\
\hline

\end{tabularx}

\end{table}

The second metric, $|\mu_{mdn} ^{i} - \theta_{true} ^{i}| < |\mu_{kde} ^{i} - \theta_{true} ^{i}|$, determines how often our proposed methodology produces superior estimates for individual parameters in a free parameter set. In some situations, one might find that the estimates obtained for a subset of the free parameters by the method of \citet{Grazzini_et_al_2017} are superior, even if the overall estimate for the entire free parameter set is not as good. Nevertheless, we find that in over $80\%$ of cases, our methodology also results in superior estimates at the level of individual parameters, a comfortable majority. It should also be noted that in virtually all situations where $|\mu_{mdn} ^{i} - \theta_{true} ^{i}| > |\mu_{kde} ^{i} - \theta_{true} ^{i}|$, such as some cases of $b_2$ and $b_3$ in the \citet{Brock_Hommes_1998} model, and $\sigma_1$ and $\sigma_2$ in the random walk model, the differences in the estimates produced by both methods are incredibly small. In contrast to this, a sizeable number of cases where $|\mu_{mdn} ^{i} - \theta_{true} ^{i}| < |\mu_{kde} ^{i} - \theta_{true} ^{i}|$, such as $g_2$ and $g_3$ in the \citet{Brock_Hommes_1998} model, and $\eta$ in the \citet{Franke_Westerhoff_2012} model, are characterised by comparatively large differences in the estimates obtained by the competing approaches. This suggests that our proposed methodology also demonstrates a degree of dominance at the level of individual parameters.

The final metric, $\sigma_{mdn} ^{i} < \sigma_{kde} ^{i}$, indicates how frequently our proposed methodology results in reduced posterior standard deviations for individual parameters, which occurs in slightly below $80\%$ of the considered cases, again a comfortable majority\footnote{On closer inspection, it appears that our methodology results in reduced posterior standard deviations more often for parameter sets consisting of more than $2$ free parameters, which may hint at the possibility of the uncertainty of estimation increasing less rapidly for our approach than for the method of \citet{Grazzini_et_al_2017} as the number of free parameters is increased. Ultimately, further investigation would be required to verify this hypothesis.}.

Based  on the evidence presented by the above metrics as a whole, it would appear that our proposed methodology does indeed compare favourably to that of \citet{Grazzini_et_al_2017}, which was itself already shown to dominate a number of other contemporary approaches in the literature by \citet{Platt_2019}. This ultimately validates our method as a worthwhile addition to the growing toolbox of estimation methods for economic simulation models.

\section{Practical Considerations}

\subsection{Choosing the Lag Length} \label{Lag_Robustness_Demo}

As previously stated, we set $L = 3$ in all estimation experiments involving our proposed method. Naturally, one may wonder whether this is an arbitrary choice or if there is a systematic way of choosing $L$. Similarly, one may also wonder if the obtained results are robust to this choice, even if only to some extent. We now address both issues.

\begin{figure}[H]

\centering

\includegraphics[width=1\linewidth]{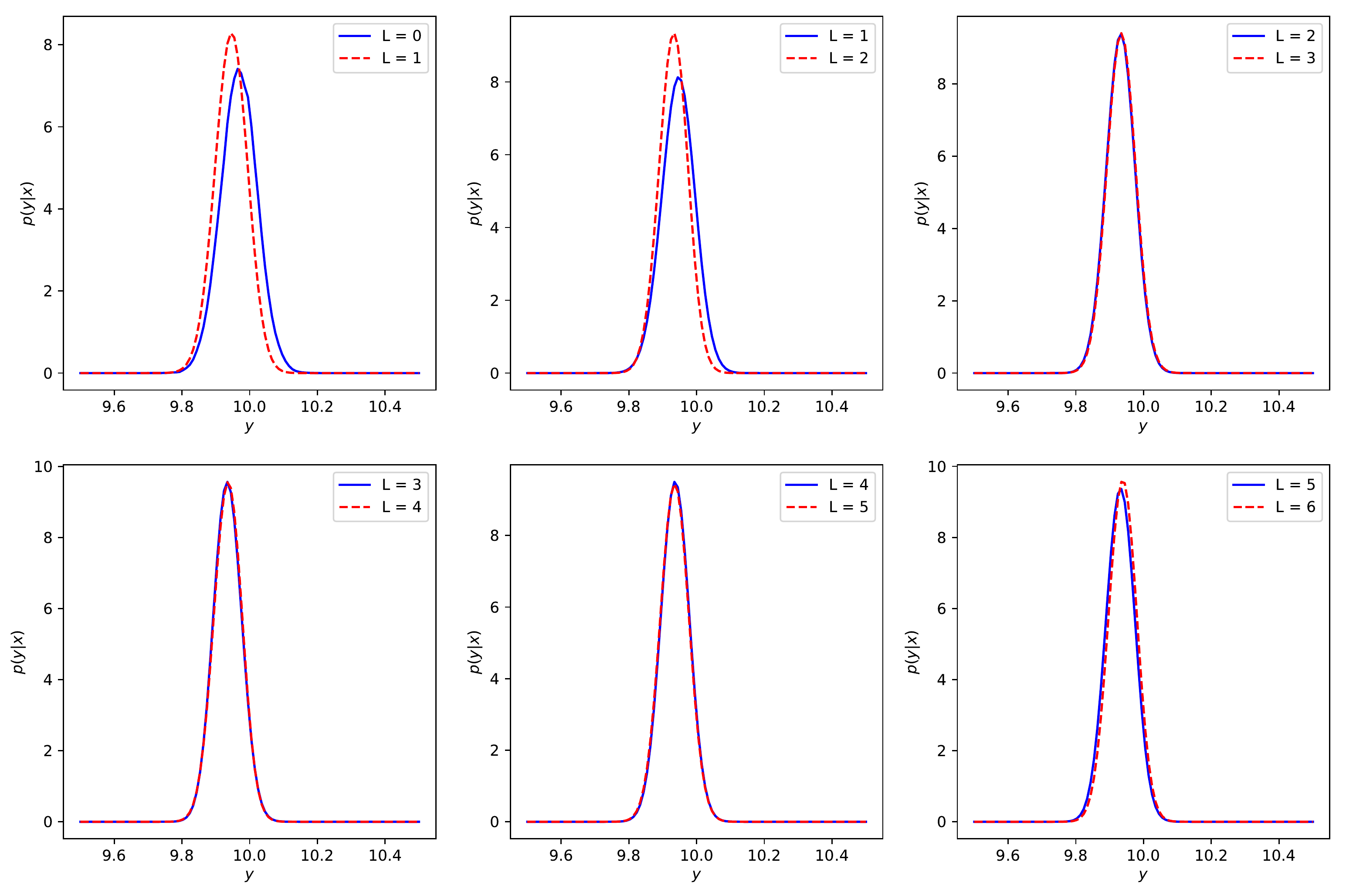}

\caption{A demonstration of the sensitivity of the conditional density estimates to the choice of lag length for a typical example of the \citet{Brock_Hommes_1998} model.}  \label{BH_Robustness}

\end{figure}

When applying the proposed methodology, we observed a phenomenon that appeared to be relatively consistent throughout the experiments. In more detail, we observe that while increasing $L$ initially has a pronounced effect on the estimated conditional densities, there exists some $L^{*} \geq 0$ such that for $L \geq L^{*}$,
\begin{equation}
p \left(\bm{x}_{t, i} ^{sim} \big| \bm{x}_{t - L, i} ^{sim}, \dots, \bm{x}_{t - 1, i} ^{sim}: \bm{\theta}\right) \simeq p \left(\bm{x}_{t, i} ^{sim} \big| \bm{x}_{t - L - 1, i} ^{sim}, \dots, \bm{x}_{t - 1, i} ^{sim}: \bm{\theta}\right),
\end{equation}
or, in other words, the MDN essentially ignores the additional lags.

\begin{figure}[H]

\centering

\includegraphics[width=1\linewidth]{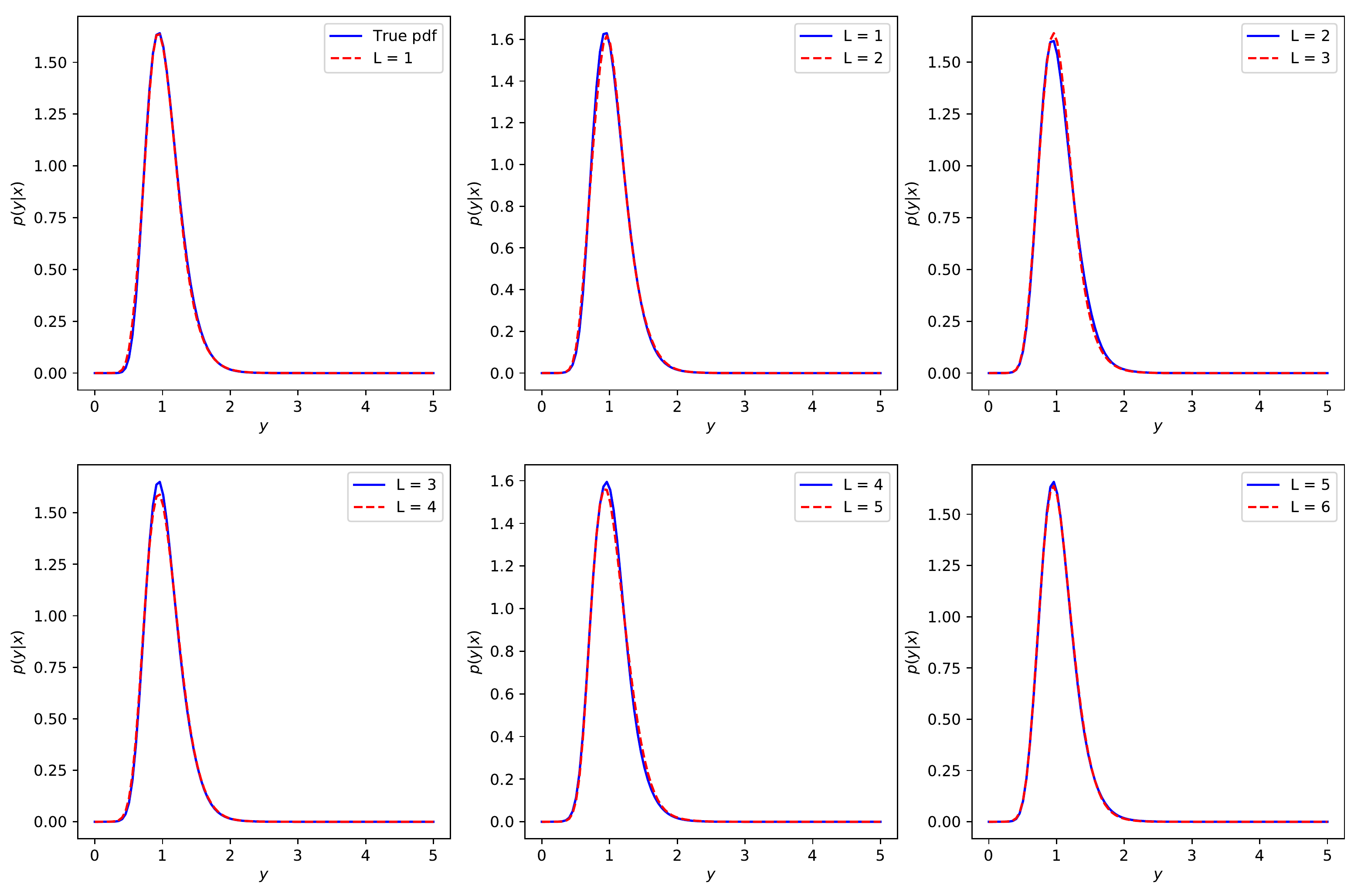}

\caption{A demonstration of the sensitivity of the conditional density estimates to the choice of lag length for i.i.d. random samples following a log-normal distribution, $LN(0, 0.25)$.} \label{LN_Robustness}

\end{figure}

\begin{figure}[H]

\centering

\includegraphics[width=1\linewidth]{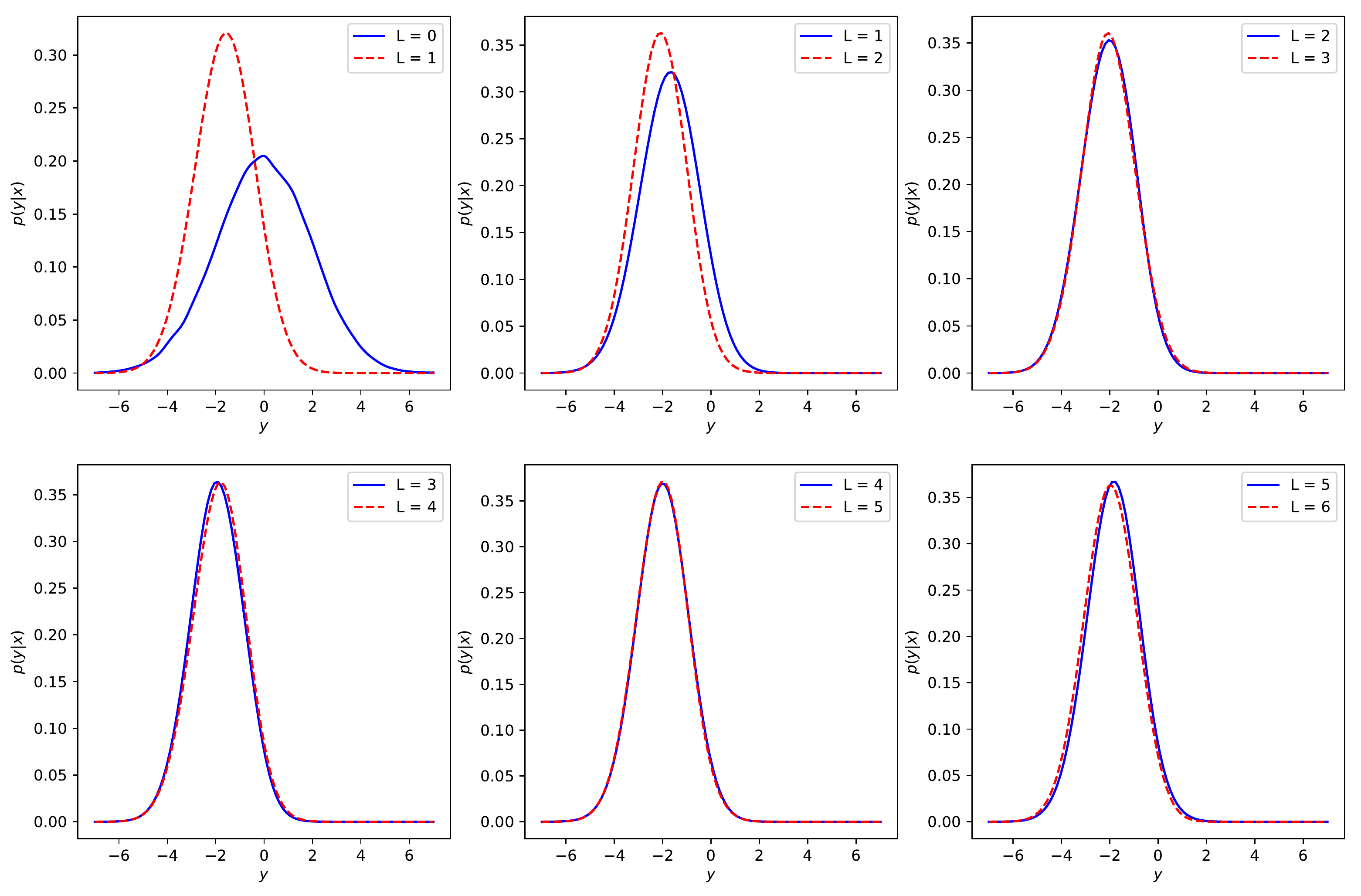}

\caption{A demonstration of the sensitivity of the conditional density estimates to the choice of lag length for an AR($2$) model, $x_{t + 1} = 0.45 x_t + 0.45 x_{t - 1} + \epsilon_t$, where $\epsilon_t \sim \mathcal{N}(0, 1)$.} \label{AR2_Robustness}

\end{figure}

We illustrate this graphically in Figure \ref{BH_Robustness}. Here, we train an MDN on $100$ realisations of length $1000$ generated using the \citet{Brock_Hommes_1998} model initialised using parameter set $1$. We then randomly draw an arbitrary sequence of $6$ consecutive values from a time series of length $1000$, also generated by the \citet{Brock_Hommes_1998} model. This then allows us to use the MDN to plot the conditional density functions for differing choices of $L$, conditioned on the values generated in the previous step, and observe the aforementioned trend.

Repeating this exercise on models for which the true lag, $L_{true}$, is known a-priori (see Figures \ref{LN_Robustness} and \ref{AR2_Robustness}), we see that $L^{*} = L_{true}$. This has a number of important implications. Firstly, it implies that plots of the type we have constructed here can be used as a means to systematically inform the choice of $L$ for arbitrary models. Secondly, and perhaps more importantly, it implies that if $L \geq L_{true}$, the procedure should demonstrate at least some robustness to the choice of lag, provided that the MDN is sufficiently expressive and sufficiently well-trained. This explains why simply setting $L = 3$ resulted in a high level of estimation performance in our experiments, regardless of the considered model, since the models considered are not characterised by long-range dependencies\footnote{The interested reader should refer to Appendix \ref{Robustness_Tests} for additional discussions.}.

\subsection{Computational Costs}

At this point, one may ask whether the proposed estimation routine compares favourably to other contemporary alternatives in terms of computational costs. As stated by \citet{Grazzini_et_al_2017}, the cost of generating simulated data using a candidate model is generally dominant, particularly for large-scale models that may need to be run for several minutes in order to generate a single realisation. It is therefore imperative that any estimation methodology keep the simulated ensemble size, which we call $R$, to a minimum. 

As previously stated, we have selected $R = 100$, which results in a relatively large training set of $R(T_{sim} - L) = 99700$ training examples. This compares favourably to most alternatives in the literature on a number of grounds. Firstly, most studies which have attempted to estimate models of similar complexity make use of ensembles consisting of a far greater number of realisations, typically in excess of $R = 1000$ \citep{Barde_2017, Lamperti_2017, Lux_2018}. Secondly, the training set associated with $R = 100$ is already large relative to the complexity of the network architecture we employ\footnote{See Appendix \ref{Network_Architecture}.}.

To illustrate this point, we repeat the experiments associated with parameter set $1$ of the \citet{Brock_Hommes_1998} model, changing only the simulated ensemble size, which has been halved to $R = 50$. We find that even with this drastic decrease in the number of Monte Carlo replications, the proposed methodology still performs well and results in a lower loss function value than was obtained using the method of \citet{Grazzini_et_al_2017} in the original experiments, with a ratio of $LS_{MDN} / LS_{KDE} = 0.7249$\footnote{Here $LS_{KDE}$ is determined from the results of the original experiment involving the method of \citet{Grazzini_et_al_2017} with $R = 100$, while $LS_{MDN}$ is determined from the results of the supplementary experiment involving our proposed methodology with $R = 50$.}. This provides some evidence that even for greatly reduced ensemble sizes, our approach remains viable, and implies that the complexity of the candidate model and hence the employed neural network would likely need to be increased substantially before any increase in $R$ beyond $100$ is required.

\begin{figure}[H]

\centering

\makebox[\linewidth][c]{%
\begin{subfigure}{0.55\linewidth}
	\centering
	\includegraphics[width=1\linewidth]{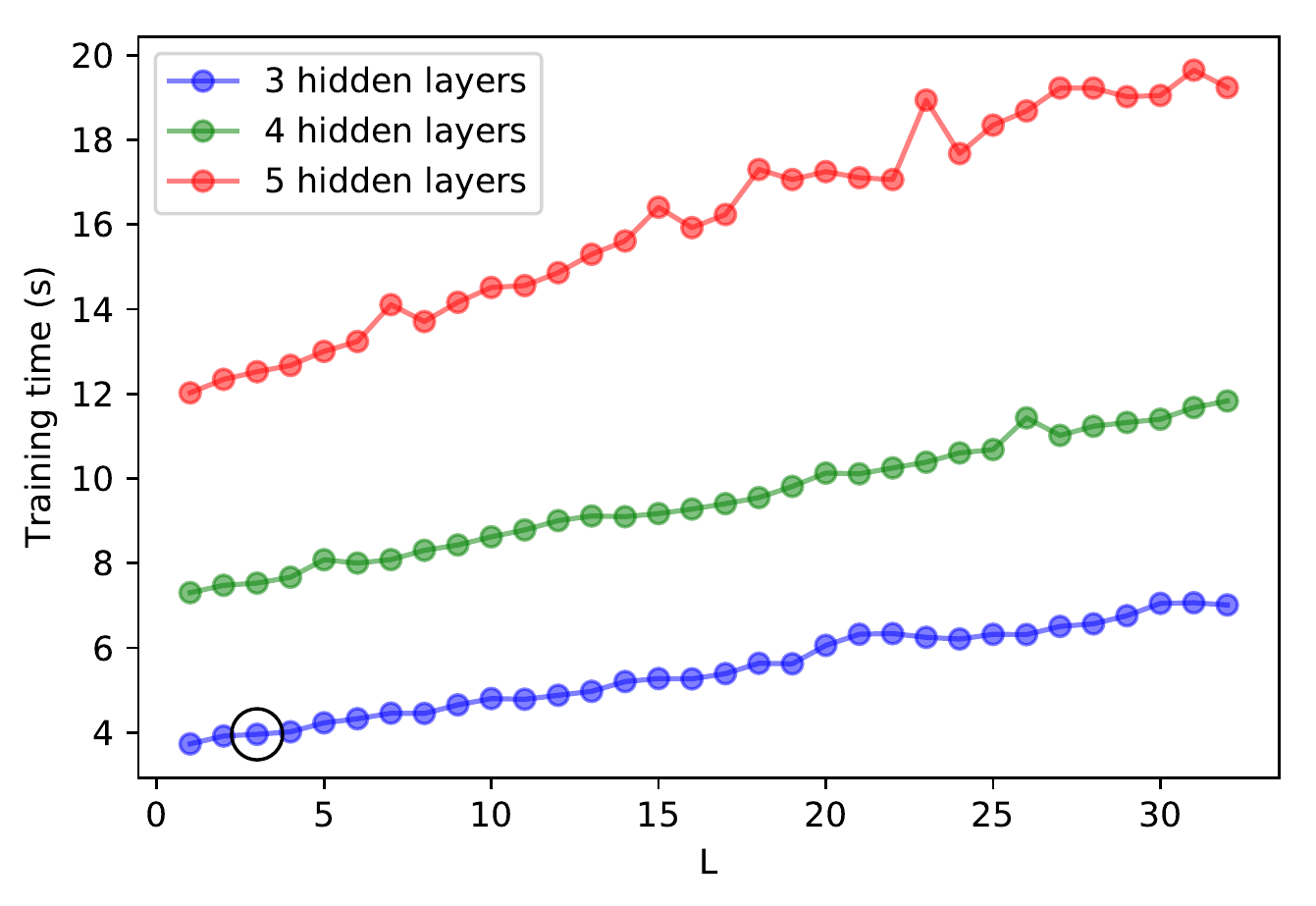}
\end{subfigure}
\begin{subfigure}{0.55\linewidth}
	\centering
	\includegraphics[width=1\linewidth]{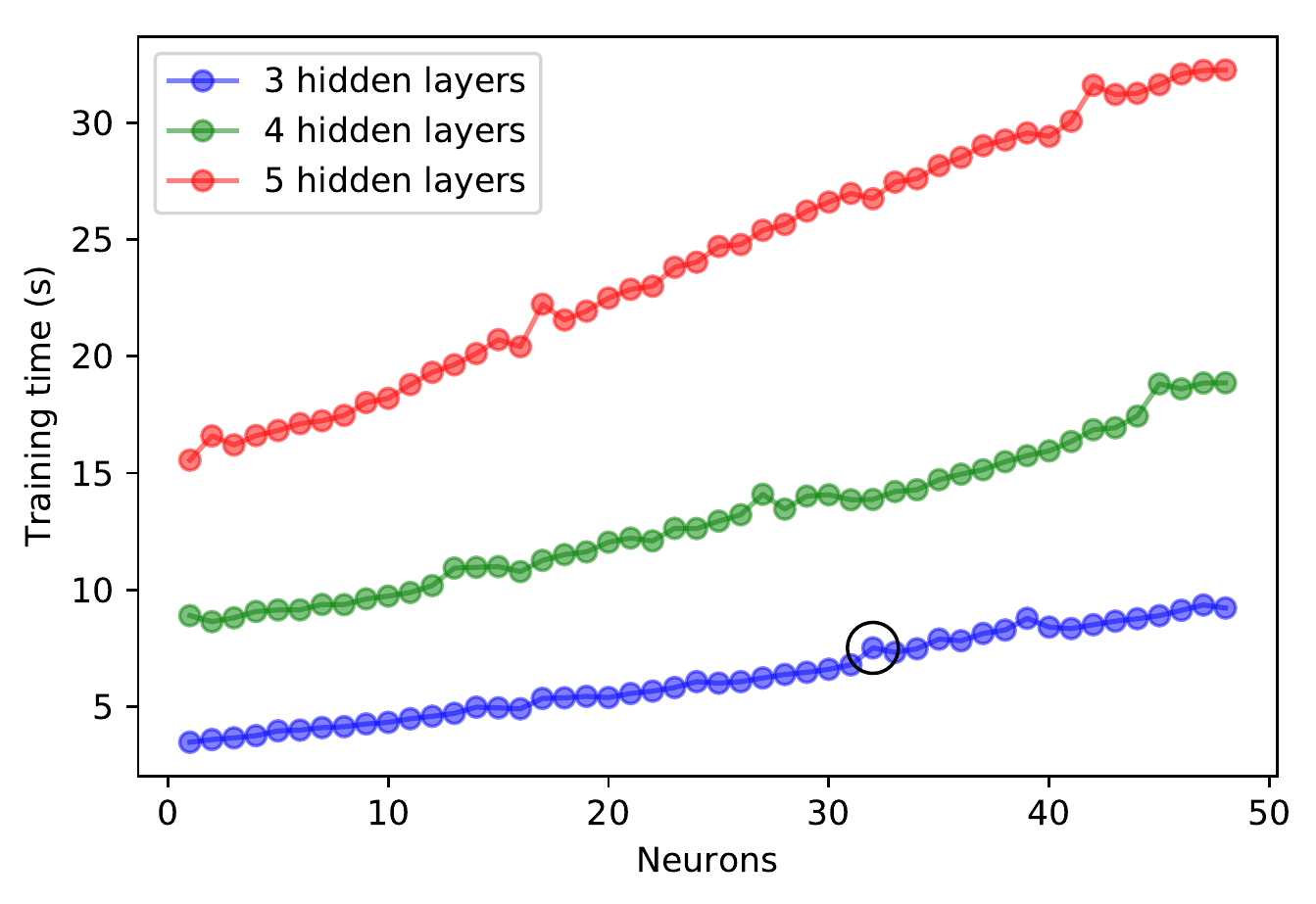}
\end{subfigure}
}

\caption{Training time for various MDN configurations on an ensemble of $100$ realisations of length $1000$ generated using the \citet{Brock_Hommes_1998} model initialised using parameter set $1$. The point indicated on both the left and right panels corresponds to the configuration employed in our estimation experiments.} \label{Computational_Cost_Plots}

\end{figure}

In addition to concerns related to the size of the simulated ensemble, it is also worthwhile to consider the actual computational costs of the neural network training procedure relative to those associated with the generation of a single model realisation. For this reason, Figure \ref{Computational_Cost_Plots} demonstrates the total training time required by various neural network configurations, most of which are larger than that of the network employed in this investigation, which typically takes $\sim5$ seconds to be completely trained. We find that even for substantially more complex neural networks than those considered in our investigation, the overall training time is still typically less than $40$ seconds, which compares favourably to the simulation time of large-scale models, and we additionally find that the increase in computational time is linear for both increases in the lag length and network width.

Further, it should be noted that GPU parallelisation was not employed when generating the aforementioned computational cost diagrams. Given the significant speedup that could be expected with the use of such hardware, typically in the region of $20\times$ \citep{Oh_Jung_2004}, we find there to be at least some evidence that the time taken to train the neural network will generally be negligible in comparison to the time taken to generate a single model realisation, even for far more sophisticated neural networks and candidate models. This would, however, require further testing that is beyond the scope of this investigation and we thus suggest that the proposed routine be applied to more sophisticated models in future work.

\section{Conclusion}

In the preceding sections, we have introduced a neural network-based protocol for the Bayesian estimation of economic simulation models (with a particular focus on ABMs) and demonstrated its estimation capabilities relative to a leading method in the existing literature. 

Overall, we find that our method delivers compelling performance in a number of scenarios, including the estimation of heterogeneous agent models typically used to test estimation procedures, and less orthodox examples, such as identifying dynamic shifts in data generated by a random walk model. In all of the cases tested, we find that our proposed methodology produces estimates closer to known ground truth values than the approach proposed by \citet{Grazzini_et_al_2017} and also find that it typically results in narrower and more sharply peaked posteriors for larger free parameter sets.

In addition to our primary findings, we also discuss practical issues related to the applicability of the proposed routine. We demonstrate that the lag length, which can be viewed as our approach's primary hyperparameter, can be systematically chosen and that the overall estimation performance demonstrates at least some robustness to this choice. Further, we provide a number of arguments as to the protocol's computational efficiency relative to a number of prominent alternatives in the literature and therefore suggest that attempts be made to apply it to models of a larger scale in future research.

\section*{Acknowledgements}

The author would like to thank J. Doyne Farmer for helpful discussions that greatly aided the process of preparing this manuscript and the UK government for the award of a Commonwealth Scholarship. Responsibility for the conclusions herein lies entirely with the author.

%----------------------------------------------------------------------------------------

% Bibliography Style
\bibliographystyle{abbrvnat}

% Bibliography File
\bibliography{DPhil_Bibliography}

%----------------------------------------------------------------------------------------

\appendix

\section{Technical Details of the Proposed Estimation Procedure \label{Method_Implementation}}

While we presented an overview of our estimation procedure in Section \ref{Experimental_Procedures}, the associated discussions were primarily illustrative and omitted several key details. We thus provide a more technical, step-by-step discussion of our approach in this section.

\subsection{Training Set Construction}

The primary aim of our methodology is the construction of an approximation to the likelihood function for a given set of parameter values, $\bm{\theta}$. In order to facilitate this process, we make the simplifying assumption that $\bm{x} ^{sim} _{t, i}(\bm{\theta})$ depends only on $\bm{x} ^{sim} _{t - L, i}(\bm{\theta}), \dots, \bm{x} ^{sim} _{t - 1, i}(\bm{\theta})$, for all $L < t \leq T$. Our problem therefore reduces to the estimation of conditional densities of the form $p \left(\bm{x}_{t, i} ^{sim} \big| \bm{x}_{t - L, i} ^{sim}, \dots, \bm{x}_{t - 1, i} ^{sim}: \bm{\theta}\right)$. 

In order to estimate the above conditional densities, we will require an appropriate dataset, which is constructed in a number of stages. The first of these stages involves the use of the candidate model to generate an ensemble of $R$ Monte Carlo replications, $\bm{X} ^{sim} (\bm{\theta}, T ^{sim}, i), i = i_0, i_0 + 1, \dots, i_0 + R - 1$, for a given value of $\bm{\theta}$. This is then followed by the construction of two ordered sets for each Monte Carlo replication $i$ in the ensemble, 
\begin{equation}
\begin{split}
\bm{X} ^{train}_i(\bm{\theta}) = \Big\{&\left\{\bm{x} ^{sim} _{1, i}(\bm{\theta}), \dots, \bm{x} ^{sim} _{L, i}(\bm{\theta})\right\}, \left\{\bm{x} ^{sim} _{2, i}(\bm{\theta}), \dots, \bm{x} ^{sim} _{L + 1, i}(\bm{\theta})\right\}, \dots, \\
&\left\{\bm{x} ^{sim} _{T - L, i}(\bm{\theta}), \dots, \bm{x} ^{sim} _{T - 1, i}(\bm{\theta})\right\}\Big\},
\end{split}
\end{equation}
and
\begin{equation}
\bm{Y} ^{train}_i(\bm{\theta}) = \left\{\bm{x} ^{sim} _{L + 1, i}(\bm{\theta}), \bm{x} ^{sim} _{L + 2, i}(\bm{\theta}), \dots, \bm{x} ^{sim} _{T, i}(\bm{\theta})\right\}.
\end{equation}
Finally, the sets $\bm{X} ^{train}_i(\bm{\theta}), i = i_0, i_0 + 1, \dots, i_0 + R - 1$ are concatenated, in order, to produce a single, larger ordered set, $\bm{X} ^{train}(\bm{\theta})$, with an analogous procedure being applied to $\bm{Y} ^{train}_i(\bm{\theta})$ to yield $\bm{Y} ^{train}(\bm{\theta})$.

In essence, $\bm{X} ^{train}(\bm{\theta})$ consists of rolling windows of length $L$ drawn from the ensemble of Monte Carlo replications, while $\bm{Y} ^{train}(\bm{\theta})$ consists of the $\bm{x} ^{sim} _{t, i}(\bm{\theta})$ values that directly follow each window in $\bm{X} ^{train}(\bm{\theta})$. Together, they form a training set of size $R(T - L)$ that can be used to approximate the required conditional densities.

\subsection{Neural Network Specification and Training}

With an appropriate dataset now constructed, we proceed with a more detailed discussion of the MDN itself. 

As a starting point, let $H$ be a feedforward neural network with input layer $\bm{h}_0$ (taking in windows of length $L$), hidden layers $\bm{h}_1, \bm{h}_2, \dots, \bm{h}_{n - 1}$, output layer $\bm{h}_{n}$, and weights and biases $\bm{\psi}$. The mixture parameters are then defined as
\begin{equation} \label{Mixture_Weights}
\bm{\alpha} = softmax(\bm{W}_{\alpha}\bm{h}_{n} + \bm{b}_{\alpha}),
\end{equation}
\begin{equation}
\bm{\mu}_{k} = \bm{W}_{\mu_k}\bm{h}_{n} + \bm{b}_{\mu_k},
\end{equation}
and
\begin{equation} \label{Cov_Mat}
\bm{\Sigma}_{k} = diag(\bm{\sigma}_{k} ^2),
\end{equation}
where $diag(\bm{x})$ is a diagonal matrix with diagonal $\bm{x}$ and
\begin{equation} \label{Log_Var}
\log \bm{\sigma} ^{2} _{k} = \bm{W}_{\sigma_k}\bm{h}_{n} + \bm{b}_{\sigma_k}.
\end{equation}
This results in an expanded neural network with weights and biases 
\begin{equation}
\bm{\phi} = \left\{\bm{\psi}, \bm{W}_{\alpha}, \bm{b}_{\alpha}, \bm{W}_{\mu_k}, \bm{b}_{\mu_k}, \bm{W}_{\sigma_k}, \bm{b}_{\sigma_k}\right\}
\end{equation}
that takes windows of length $L$ as input and outputs $\bm{\alpha}$, $\bm{\mu}_{k}$, and $\bm{\Sigma}_{k}$ as defined above.

At this stage, there are a number of nuances worth highlighting. In Eqn. \ref{Mixture_Weights}, notice that we make use of the $softmax$ function. This ensures that the mixture weights, $\bm{\alpha}$, are strictly positive and sum to one, as required. Additionally, notice that in Eqn. \ref{Cov_Mat} we consider a diagonal rather than a full covariance matrix\footnote{It should be noted that the universal density approximation properties of Gaussian mixtures still apply for diagonal covariance matrices.}. If we had not made such an assumption, we would have to ensure that the covariance matrices returned by our neural network were positive definite. Though possible in principle, this would significantly increase the number of network parameters and have a potentially detrimental effect on computational performance \citep{Rothfuss_et_al_2019}. Finally, it should be apparent from Eqn. \ref{Log_Var} that the neural network outputs a vector of log variances rather than the diagonal covariance matrix, allowing us to avoid imposing positivity constraints on the network output.

Now, all that remains is the training of our constructed network, which is achieved through the application of maximum likelihood estimation to our training set. Denoting by $\bm{X}_m ^{train}$ the $m$-th entry in $\bm{X} ^{train}(\bm{\theta})$ (with $\bm{Y}_m ^{train}$ being similarly defined), maximum likelihood estimation is equivalent to solving
\begin{equation}
\arg\min_{\bm{\phi}} -\sum_{m = 1} ^{R (T - L)} \log \sum_{k = 1} ^{K} \alpha_k \left(\bm{X}_m ^{train}\right) \mathcal{N}\left(\bm{Y}_m ^{train} \big| \bm{\mu}_k\left(\bm{X}_m ^{train}\right), \bm{\Sigma}_k\left(\bm{X}_m ^{train}\right)\right)
\end{equation}
using stochastic gradient descent methods.

\subsection{Data Normalisation and Regularisation}

While the scheme we have just described could be applied as is, it is likely to perform suboptimally in its current form. This is because neural networks, like most machine learning techniques with a large number of free parameters, have a tendency to overfit the training data and thus perform poorly out-of-sample, particularly when the training set is small \citep{Murphy_2012}. In practice, this is often addressed using early stopping, a technique that requires a percentage of the data to be kept separate from the training set in order to evaluate out-of-sample performance during each epoch \citep{Prechelt_1998}. Such a solution is, however, undesirable in our context, since it requires the generation of additional data, an expensive undertaking for large-scale simulation models.

Fortunately, \citet{Rothfuss_et_al_2019} present a set of best practices for conditional density estimation using neural networks that provides an alternative solution for overfitting. In particular, a technique called noise regularisation is employed, in which small random perturbations are applied to the data during the training process. It can be shown that this ultimately results in a complexity penalty that favours smoother density estimates that are less prone to overfitting \citep{Rothfuss_et_al_2019}. For this reason, we apply Gaussian perturbations to training examples in $\bm{X} ^{train}(\bm{\theta})$ and $\bm{Y} ^{train}(\bm{\theta})$, which we denote by
\begin{equation}
\bm{\xi}_{x} \sim\mathcal{N}(0, \eta_x\bm{I}) \text{ and } \bm{\xi}_{y} \sim \mathcal{N}(0, \eta_y\bm{I}),
\end{equation}
respectively. 

It should be apparent that the degree of regularisation depends directly on the magnitudes of the standard deviations $\eta_x$ and $\eta_y$ relative to the range of variation in the training data\footnote{As an example, setting $\eta_x = 0.5$ would result in a substantial amount of regularisation for training examples that take values in $[0, 1]$, while essentially having no effect for training examples taking values in $[0, 1000]$.}. This implies that $\eta_x$ and $\eta_y$ would have to be adjusted for each candidate model in order to result in the same degree of regularisation. \citet{Rothfuss_et_al_2019} therefore propose a data normalisation scheme that ensures the training data exhibits zero mean and unit variance, eliminating the need to retune these hyperparameters for each candidate model. This is achieved through the application of a simple transformation to each training example.

Letting $\hat{\bm{\mu}}_x$ and $\hat{\bm{\sigma}}_x$ be vectors that contain estimates of the mean and standard deviation along each dimension for training examples in $\bm{X} ^{train}(\bm{\theta})$, this transformation is given by
\begin{equation} \label{Transformation}
\tilde{\bm{X}}_m ^{train} = diag(\hat{\bm{\sigma}}_x) ^{-1} (\bm{X}_m ^{train} - \hat{\bm{\mu}}_x),
\end{equation}
with $\hat{\bm{\mu}}_y$, $\hat{\bm{\sigma}}_y$ and $\tilde{\bm{Y}}_m ^{train}$ being defined analogously.

Once the network has been trained on the normalised dataset, we are required to evaluate $\tilde{f}(\bm{x}, \bm{y}, \bm{\phi})$, originally defined in Eqn. \ref{Network_Inference}. This is achieved through a simple procedure. Firstly, the normalisation transform is applied to $\bm{x}$ and $\bm{y}$ using the same $\hat{\bm{\mu}}_y$, $\hat{\bm{\sigma}}_y$, $\hat{\bm{\mu}}_x$ and $\hat{\bm{\sigma}}_x$ values defined in Eqn. \ref{Transformation}, yielding $\tilde{\bm{x}}$ and $\tilde{\bm{y}}$. $\tilde{\bm{x}}$ is then fed through the trained neural network to yield corresponding mixture parameters, allowing us to evaluate the density at $\tilde{\bm{y}}$, which we denote by $\tilde{g}(\tilde{\bm{x}}, \tilde{\bm{y}}, \tilde{\bm{\phi}})$. It should be noted that $\tilde{g}$ does not directly correspond to $\tilde{f}$, since we have made a change of variables and the volume of the probability density is not preserved under the normalisation transform for $\hat{\bm{\sigma}}_y \neq 1$. \citet{Rothfuss_et_al_2019} do, however, prove that
\begin{equation}
\tilde{f}(\bm{x}, \bm{y}, \bm{\phi}) = \frac{1}{\prod_{j = 1} ^{J} \hat{\sigma}_y ^{(j)}} \tilde{g}(\tilde{\bm{x}}, \tilde{\bm{y}}, \tilde{\bm{\phi}}),
\end{equation}
where $\hat{\sigma}_y ^{(j)}$ is the $j$-th element of $\hat{\bm{\sigma}}_y$, allowing us to easily calculate the required density.

\subsection{Neural Network Architecture} \label{Network_Architecture}

In essence, we have defined a general neural network-based approach to simulation model estimation that is independent of the specific network architecture (number of hidden layers, number of neurons, type of activation functions, and so on) used. Nevertheless, for the sake of completeness, we briefly introduce the (relatively simple) architecture employed in our study, which is used consistently throughout unless stated otherwise.

For the mixture model itself, we set the number of mixture components to be $K = 16$, with the associated mixture parameter network consisting of $3$ hidden layers, each with $32$ neurons and ReLU activations. This was trained using the well-known Adam optimiser \citep{Kingma_Ba_2015} over $12$ epochs\footnote{Any improvements in the likelihood for subsequent epochs were generally negligible.}, with a batch size of $512$ and noise regularisation parameters $\eta_x = \eta_y = 0.2$.

The above architecture, which performed well for all of the estimation tasks conducted, was, perhaps rather surprisingly, the first architecture we considered and was chosen by hand rather than through an automated optimisation procedure. Attempts to improve performance by increasing the number of hidden layers, neurons, and mixture components seemed to have little effect, suggesting that the proposed network is sufficiently expressive to produce high-quality density estimates for our considered set of problems. We suspect that this will likely hold for other models of similar complexity and therefore make the recommendation that our proposed architecture be used as a baseline for future investigations employing this estimation methodology.

For more complex models, however, it may be necessary to construct more expressive networks and in such cases we would suggest that some form of hyperparameter optimisation be carried out. This is beyond the scope of our investigation, however, and we thus leave it to future research.

\section{Technical Details of the Employed Sampling Strategy} \label{MCMC_Alg}

In this section, we briefly discuss the adaptive Metropolis-Hastings algorithm that has been employed in all of the conducted estimation experiments. Our discussion here is mainly illustrative and positioned in the context of our investigation. The interested reader should therefore refer to the original contribution by \citet{Griffin_Walker_2013} for theoretical justifications and a more general discussion.

In essence, the approach is centred on the idea of maintaining a set of samples, $\bm{\theta}_s = \left\{\bm{\theta_{s}} ^{(1)}, \bm{\theta_{s}} ^{(2)}, \dots,  \bm{\theta_{s}} ^{(N)} \right\}, s = 1, 2, \dots, S$, that is updated for a desired number of iterations. Initially, the set consists of samples drawn uniformly from the space of feasible parameter values, $\bm{\Theta}$, but eventually converges to be distributed according to $p(\bm{\theta}|\bm{X})$. This is achieved through the construction of an adaptive proposal distribution that is dependent on the current samples, $\bm{\theta}_s$, which can be summarised algorithmically as follows:
\begin{enumerate}
\item Sample $\bm{z}$ according to $\tilde{p}\left(\bm{z} \big| \bm{\theta_{s}} ^{(1)}, \bm{\theta_{s}} ^{(2)}, \dots,  \bm{\theta_{s}} ^{(N)} \right)$, which is determined by applying KDE to $\bm{\theta_{s}} ^{(1)}, \bm{\theta_{s}} ^{(2)}, \dots,  \bm{\theta_{s}} ^{(N)}$.
\item Propose the switch of $\bm{z}$ with $\bm{\theta_{s}} ^{(n)}$, where $n$ is chosen uniformly from $\left\{1, 2, \dots, N \right\}$.
\item Accept the switch with probability
\begin{equation}
\alpha = \min \left\{1, \frac{p\left(\bm{z} \big| \bm{X}\right) \tilde{p}\left(\bm{\theta_{s}} ^{(n)} | \bm{\theta_{s}} ^{(1)}, \bm{\theta_{s}} ^{(2)}, \dots,  \bm{\theta_{s}} ^{(n - 1)}, \bm{z},  \bm{\theta_{s}} ^{(n + 1)}, \dots,  \bm{\theta_{s}} ^{(N)}\right)}{p\left(\bm{\theta_{s}} ^{(n)} | \bm{X}\right) \tilde{p}\left(\bm{z} | \bm{\theta_{s}} ^{(1)}, \bm{\theta_{s}} ^{(2)}, \dots,  \bm{\theta_{s}} ^{(N)}\right)} \right\}.
\end{equation}
\item If accepted, set $\bm{\theta}_{s + 1} = \bm{\theta}_{s}$ with $\bm{\theta_{s}} ^{(n)}$ replaced by $\bm{z}$, otherwise simply set $\bm{\theta}_{s + 1} = \bm{\theta}_{s}$.
\end{enumerate}

Repeating the above for $S$ iterations, we obtain a sequence of sample sets that can be used to compute expectations of the form
\begin{equation}
\mathbb{E}\left[g(\bm{\theta})\right] = \frac{1}{NS} \sum_{s = 1} ^{S} \sum_{n = 1} ^{N} g\left(\bm{\theta_{s}} ^{(n)}\right).
\end{equation}
 
In our investigation, we set $S = 5000$ and $N = 70$ in all cases, with convergence typically observed at some point before $s = 1500$, leading us to discard the first $1500$ sets as part of a burn-in period. When constructing the posterior samples, we repeat this entire sampling process $5$ times and collect the obtained sets to form a larger collection of $5 \times 3500 \times 70 = 1225000$ samples\footnote{Note that since we only update a single sample during each step, the Monte Carlo variance still decreases at the standard rate of $\frac{1}{\sqrt{S}}$.}.
 
Ultimately, this has become our MCMC algorithm of choice for two main reasons:
\begin{enumerate}
\item The number of iterations required to reach convergence in random walk Metropolis-Hastings algorithms depends significantly on the initialisation of the algorithm. If, for example, the initial candidate parameter set has a particularly low posterior density, it could take a substantial period of time before convergence is observed. Since the algorithm proposed by \citet{Griffin_Walker_2013} is initialised using a sample of points from a number of areas of the parameter space, this problem is less pronounced.
\item Most random walk Metropolis-Hastings algorithms require careful tuning of the proposal distribution, usually with the aim of obtaining an acceptance rate of roughly $25\%$, in order to ensure a good balance between local exploration of high density areas of the parameter space and global coverage of the parameter space as a whole \citep{Robert_Casella_2010}. This can be difficult to achieve in practice, making an adaptive approach that determines the proposal distribution automatically particularly appealing.
\end{enumerate}
 
\section{Robustness Tests} \label{Robustness_Tests}

In Section \ref{Lag_Robustness_Demo}, we provided evidence that our proposed estimation procedure demonstrates some robustness relative to the choice of lag length, $L$. Here, we provide a more complete demonstration by repeating all of the previously conducted estimation experiments involving our approach, changing only the lag length, which we have increased to $L = 4$. Referring to the summary presented in Table \ref{Overall_Summary_Robustness}, we find that the overall performance of the procedure relative to our chosen benchmark is virtually unchanged\footnote{Since there are a total of $27$ individual parameter cases, the percentage shifts correspond to changes in only a single binary relation for both $|\mu_{mdn} ^{i} - \theta_{true} ^{i}| < |\mu_{kde} ^{i} - \theta_{true} ^{i}|$ and $\sigma_{mdn} ^{i} < \sigma_{kde} ^{i}$.}, verifying the robustness of our conclusions.

\begin{table}[H]

\setlength\extrarowheight{3pt}

\caption{Estimation Result Summary Across All Models for $L = 4$} \label{Overall_Summary_Robustness}

\begin{tabularx}{\linewidth}{XX}

\hline
\midrule
Outcome & Percentage of Cases \\
\midrule
\hline
$LS_{mdn} < LS_{kde}$ & $100$ \\
$|\mu_{mdn} ^{i} - \theta_{true} ^{i}| < |\mu_{kde} ^{i} - \theta_{true} ^{i}|$ & $77.78$ \\
$\sigma_{mdn} ^{i} < \sigma_{kde} ^{i}$ & $74.07$ \\
\hline

\end{tabularx}

\end{table}

\end{document}